\documentclass[acmtog,nonacm]{acmart}

\usepackage{booktabs} 
\usepackage{graphicx}
\usepackage{booktabs} 
\usepackage{subcaption}
\usepackage[percent]{overpic}
\usepackage{xcolor}
\usepackage{mathtools}
\usepackage{xspace}
\usepackage{makecell}
\usepackage{wrapfig}
\usepackage[normalem]{ulem}
\usepackage{amsfonts}
\usepackage{algorithmic}
\usepackage{algorithm}
\usepackage{multirow}

\newlength{\figcol}

\acmJournal{TOG}

\begin{document}
\title{Patch-Based Generalized Neural Texture Upscaling}

\author{Shin Fujieda}
\affiliation{%
  \institution{Advanced Micro Devices, Inc.}
  \city{Tokyo}
  \country{Japan}}
\email{Shin.Fujieda@amd.com}
\orcid{0000-0002-2472-7365}

\author{Takahiro Harada}
\affiliation{%
  \institution{Advanced Micro Devices, Inc.}
  \city{Santa Clara}
  \country{USA}}
\email{Takahiro.Harada@amd.com}
\orcid{0000-0001-5158-8455}

\renewcommand\shortauthors{Shin Fujieda and Takahiro Harada}

\begin{abstract}
We propose a lightweight neural texture upscaling method. A single generalized MLP is trained on a diverse collection of texture patches and integrated into the texture sampling stage, upscaling low-resolution textures on-the-fly at rendering time. Because the model is generalized rather than overfit to specific content, no per-asset training is required - the same model applies to any existing scene without modifying texture data or the asset pipeline. Storing lower-resolution base textures reduces disk footprint, upscaling at sampling time eliminates the need to store high-resolution textures in VRAM, and sharing a single set of model weights across all textures ensures non-divergent GPU execution. We evaluate our method across multiple texture classes, demonstrating consistent improvement over analytic upscaling methods in perceptual quality. 
\end{abstract}

%
%

%
%

\begin{teaserfigure}
    \centering
    \begin{minipage}[t]{0.41\linewidth}
        \centering
        \includegraphics[width=\linewidth]{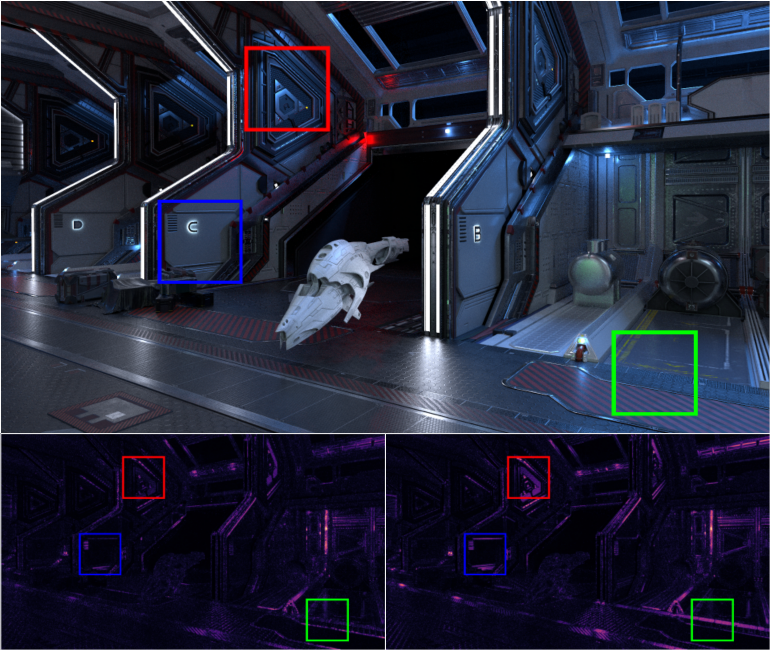}
    \end{minipage}
    \begin{minipage}[t]{0.58\linewidth}
        \centering
        \includegraphics[width=\linewidth]{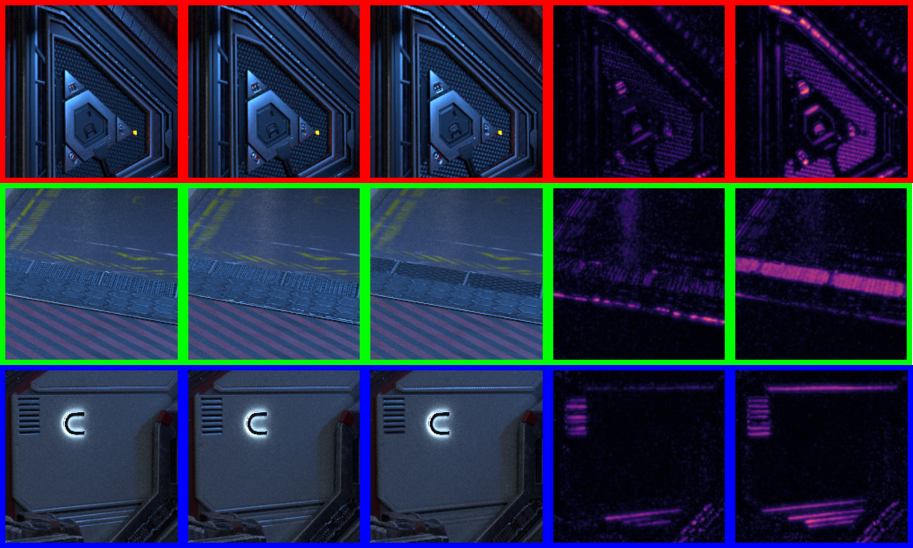}
    \end{minipage}
    \caption{Rendered image comparison using sets of textures in the \textsc{Hanger} scene, upscaled by different methods. The top-left image is the reference image with the original texture. The bottom-left images show FLIP errors for images rendered with textures upscaled using our neural texture upscaling and analytic bicubic upscaling. FLIP scores are $0.052$ and $0.057$. Zoom-ups are the reference, images rendered with textures upscaled using our neural texture upscaling and bicubic upscaling, and their FLIP images.}
    \label{fig:teaser}
\end{teaserfigure}

\newcommand{\shin}[1]{\textcolor{red}{\textbf{#1}}}
\newcommand{\takahiro}[1]{\textcolor{orange}{\textbf{#1}}}

\maketitle

\section{Introduction}

Textures and texture mappings are fundamental techniques for representing surface detail that cannot be captured by geometry alone. As display resolutions continue to increase, rendered images must achieve correspondingly higher resolutions, driving demand for higher-resolution textures. However, this comes at significant cost: higher-resolution textures consume more disk space and VRAM when rendering on a GPU. Greater disk space translates directly to larger download sizes for games and applications, while increased VRAM consumption raises hardware requirements for end users.                            

One class of approaches to this problem is offline neural image upscaling. Methods such as~\cite{dong2014learning, long2025progressive} produce high-quality upscaled images using deep neural networks (NNs) trained on large image datasets. These models are generalized: they require no training on the target content and can therefore be applied to existing assets without modifying the content creation pipeline. However, they are not designed to run at rendering time because of the computational cost and missing capability of random access, which is the typical use case for runtime. The option to use them is to perform upscaling as a preprocessing step before rendering begins. This means that the upscaled high-resolution textures must then be stored and loaded into VRAM at runtime. Consequently, while these methods can improve texture quality and reduce download size when applied to games, they offer no reduction in disk footprint or VRAM pressure.

Neural Texture Compression (NTC)~\cite{Vaidyanathan2023} takes a fundamentally different approach by representing each texture set as a compact neural network decoded on-the-fly at texture sampling time. This achieves high compression ratios, particularly when compressing correlated texture sets used in physically based rendering (PBR) such as diffuse, normal, ARM (AO, Roughness, Metallic), where correlation across channels can be exploited. Since decompression occurs at sampling time, NTC reduces both disk footprint and VRAM simultaneously. However, NTC presents practical challenges for adoption. First, because the neural network is overfit to each texture set, compression must be re-executed per asset, adding overhead to the content creation pipeline. Second, since each texture set is encoded as a separate network with different weights, concurrent sampling of multiple texture sets within a single GPU wavefront causes execution divergence — unfavorable for the matrix cores that are the key to GPU inference performance, and particularly problematic during non-primary material execution where many texture sets may be active simultaneously~\cite{AMD_CDNA2021, NVIDIA_Volta2017}.

Neural Texture Block Compression (NTBC)~\cite{Fujieda2024NTBC} combines neural network-based encoding with standard GPU-hardware block-compression formats~\cite{Akenine-Moller2018}. A neural network is used to produce block-compressed representations, while standard hardware texture units handle decoding — avoiding runtime neural inference cost and the divergence issues of NTC. However, because the output must conform to block compression constraints, the quality gains over conventional block compression are inherently bounded. Similar to NTC, NTBC must be applied per asset during content creation, adding pipeline overhead and precluding application to existing unmodified content.

We propose a lightweight neural texture upscaling method designed for ease of integration. We train a single generalized Multi-Layer Perceptron (MLP) on a diverse collection of texture patches from publicly available texture datasets. At rendering time, this model is integrated into the texture sampling stage: the renderer fetches a low-resolution texture, and the MLP upscales it on-the-fly. Because the model is generalized rather than overfit to specific content, no per-asset training is required: the same model can be applied to any texture set in any scene, making it directly compatible with existing games and rendering pipelines without modifying texture data or the asset creation workflow. Storing lower-resolution base textures reduces disk footprint, and since upscaling occurs at sampling time, high-resolution textures need never reside in VRAM. Furthermore, because all textures share a single set of model weights, GPU execution remains non-divergent regardless of how many texture sets are active simultaneously. While the compression ratio does not match NTC, we believe these practical advantages, such as generality, pipeline compatibility, and uniform GPU execution, make the proposed method compelling for real-time rendering applications such as games.
\section{Related Work}

Texture compression has been a fundamental component of real-time rendering pipelines. Block compression (BC) formats come in a wide variety depending on the number of channels and quality~\cite{Akenine-Moller2018}. As the name suggests, they are block-compressed, so they are suited for random access. However, the fixed-block structure imposes a ceiling on the compression ratio and quality, which some neural network-based methods managed to break as discussed later in this section. When a texture is viewed at close range, we need to magnify the texture content. There are a few analytic filters we can use to reconstruct signals higher than the original texture resolution. Bilinear, bicubic, and Lanczos are the popular choices for the interpolation. Bilinear interpolation uses the four nearest neighboring texels, while bicubic and Lanczos operate on a larger neighborhood, enabling higher-quality reconstruction. They are also called resampling or upsampling. We call them upsampling in the rest of this paper, as they can be viewed as upsampling operations on texture data. Although these analytic methods are computationally inexpensive, they are fundamentally limited in reconstruction quality because they cannot recover details that are absent in the low-resolution textures.

 A learning-based method or a neural network overcomes this limitation. The trend has shifted from CNN-based upscaling methods such as SRCNN~\cite{dong2014learning} to transformer-based upscaling methods such as SwinIR~\cite{liang2021swinir} and, most recently, PFT~\cite{long2025progressive}. These methods are general-purpose image upscalers that learn the image characteristics to reconstruct missing finer details. However, these methods are not designed to run in real time like real-time analytic upscaling methods. As they are computationally expensive and operate on the full image, they are not suitable to be executed at texture sampling in real-time rendering. 

Another class of research is to apply learned upsampling specifically to PBR texture content. MATUP~\cite{https://doi.org/10.1111/cgf.15151} repurposes pre-trained RGB image upsamplers to super-resolve SVBRDF maps by optimizing a local MLP filter per material, with a radiance-domain loss to ensure physically consistent upscaling. PBR-SR~\cite{chen2025pbrsr} similarly leverages pre-trained 2D image super-resolution priors and refines PBR textures through iterative differentiable rendering against multi-view constraints. Both methods produce high-quality results but require per-asset optimization, making them unsuitable for integration into real-time rendering pipelines. Our method, in contrast, is a fully generalized model trained across a diverse set of PBR textures, requiring no per-asset processing. This allows it to be applied at texture-sampling time without any content creation overhead, directly replacing the conventional analytic upsampling step in the rendering pipeline.

Neural Texture Compression (NTC) is specifically targeted to be executed in this context~\cite{Vaidyanathan2023}. Instead of using larger models such as CNNs, NTC uses a small MLP with an input encoder, which makes it possible to do local per-texel evaluation at sampling time, which CNN and transformer-based models cannot do. Another fundamental difference is that NTC is not a generalized model, but a set of PBR textures must be overfit into a single neural network. Although it achieves impressive compression, especially with the latest input encoding research~\cite{10.1145/3806062}, using NTC adds an overhead to content creation. Since the MLP execution is local, one can run a small inference for each texture sampling operation. However, the matrix cores on today's GPUs are designed to use a group of threads for matrix computation, which is a different programming pattern from SIMT execution~\cite{AMD_CDNA2021, NVIDIA_Volta2017}. This makes it difficult to execute NTC efficiently on the GPU when there are divergences in the execution, which is usually the case~\cite{10.2312:hpg.20251173}. Neural Texture Block Compression (NTBC) is a variant of NTC, which executes neural networks to create BC-format textures~\cite{Fujieda2024NTBC}. NTBC is designed that way so that it is easier to adopt to a game production. As the neural network produces BC blocks which can be parsed and decompressed using the hardware decoder, NTBC can be executed to decompress textures into BC textures on load. NTBC also requires overfitting, so it adds overhead to the content creation pipeline. 

Our method is designed to reduce these adoption barriers. First, our method is generalized, so no need to add a processing step in the content creation pipeline. The execution divergence is eliminated as our model is generalized, so we can execute the same model for different textures. It is even possible to integrate our method into existing games by replacing the texture sampler with the neural upscaler, enabling higher-quality rendering without modifying texture assets. 
This makes our method stand in a unique position compared to all the existing works.

\begin{figure}
     \centering
     \includegraphics[width=\columnwidth]{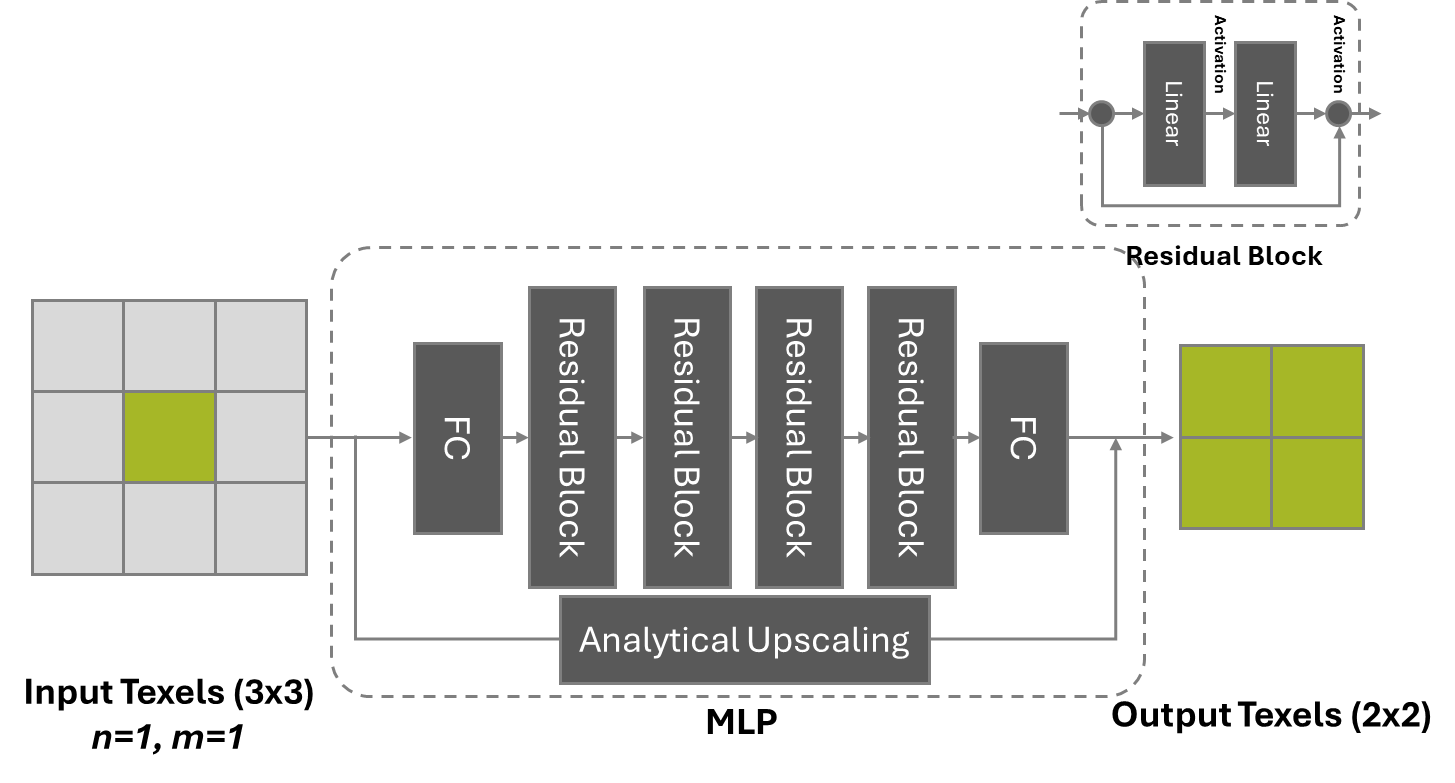}
     \caption{The network architecture of our method. In this configuration, the center texel in the input is the upscaling target. The neighbor texels are also passed to the NN. Analytically upscaled texels are added to the NN output, forming a residual connection.}
     \label{fig:diagram}
\end{figure}

\section{Method}

A key design choice of our model is locality: rather than upscaling the entire texture at once, the model upscales a small patch at texture-sampling time. Conventional texture upscaling approaches either store full high-resolution textures in GPU memory or preprocess the textures offline before rendering. Instead, our method generates high-resolution texels on demand during a texture fetch, eliminating the memory overhead of storing a full-resolution copy. Critically, operating on small patches keeps the parameter count and arithmetic cost low enough for the model to be executed directly as a GPU shader. This patch-based execution is what makes our method easier to deploy.

Given a low-resolution input patch of $n \times n$ texels, the model outputs $2n \times 2n$ texels at twice the resolution, where we call $n$ the patch size. However, upscaling a patch without surrounding context produces block artifacts at patch boundaries, because the model cannot observe the texture pattern beyond the patch edges. To address this, we extend the input patch by $m$ texels on each side. 
The network therefore receives a $(n+2m) \times (n+2m) \times c$ tensor as input and produces $2n \times 2n \times c$ texels as output, where $c$ is the number of channels. The $m$-texel margin is consumed as context and discarded from the output. In our experiments, we use $n=1$, $m=1$, and $c=3$ for RGB textures. We chose $n=1$ to minimize the number of texels fetched per sample while still producing a relatively small $2 \times 2$ output patch with a $3 \times 3$ input patch. In order to reduce the computational cost, we want to use as small an $n$ as possible. For the texels close to the texture border where the padding texels do not exist, we mirror the texture values at the boundary, assuming that textures are smooth at the boundary. 

We took an approach to upscale the texels using an analytic method first, as it is a good approximation although it is missing the high-frequency signal. Therefore, the network first upsamples the input patch to the output resolution using the analytic bilinear upscaling, which efficiently recovers the low-frequency content of the texture. The model then predicts the residual using an MLP, and the final output is the sum of the bilinear-upscaled texels and the residual. This residual formulation is effective because the bilinear baseline already provides a reasonable approximation: the MLP only needs to learn the high-frequency detail that bilinear interpolation fails to recover, which simplifies the learning problem and accelerates convergence. The input to the MLP is a flattened $(n+2m)^2 \times c$-dimensional vector. The network consists of a linear projection to a hidden dimension of $d$, followed by $L$ residual blocks. Each residual block applies two linear layers with a ReLU activation between them, adds the result to its input as a skip connection, and passes the sum through a final ReLU activation. After the residual connection, the values are clamped not to exceed $1$. The output layer then projects back to the target dimension of $2n \times 2n \times c$.  The network architecture is illustrated in Fig.~\ref{fig:diagram}.

The model is trained end-to-end using the Adam optimizer with a learning rate of $10^{-3}$. We use an L1 loss between the predicted and ground-truth high-resolution patches, which we found to produce sharper texture reconstructions than MSE. Optionally, a cosine annealing scheduler is applied to decay the learning rate to $10^{-6}$ over the course of training.



\begin{table}
  \caption{Number of textures in the datasets. } 
  \centering                                    
  \begin{tabular}{lccc}                         
  \toprule                                      
  Dataset & Diffuse & Normal & ARM \\           
  \midrule                                      
  \textsc{Poly Haven} & 565 & 565 & 563 \\
  \textsc{Hangar}     & 45 & 44 & 45 \\                      
  \textsc{Victorian}  & 37 & 36 & 36 \\                      
  \textsc{Museum}     & 56 & 29 & 30 \\                      
  \bottomrule                                   
  \end{tabular}   
  \label{tab:datasets}
\end{table}

\begin{figure*}
     \centering
     \begin{subfigure}[b]{0.42\textwidth}
         \centering
         \includegraphics[width=\textwidth]{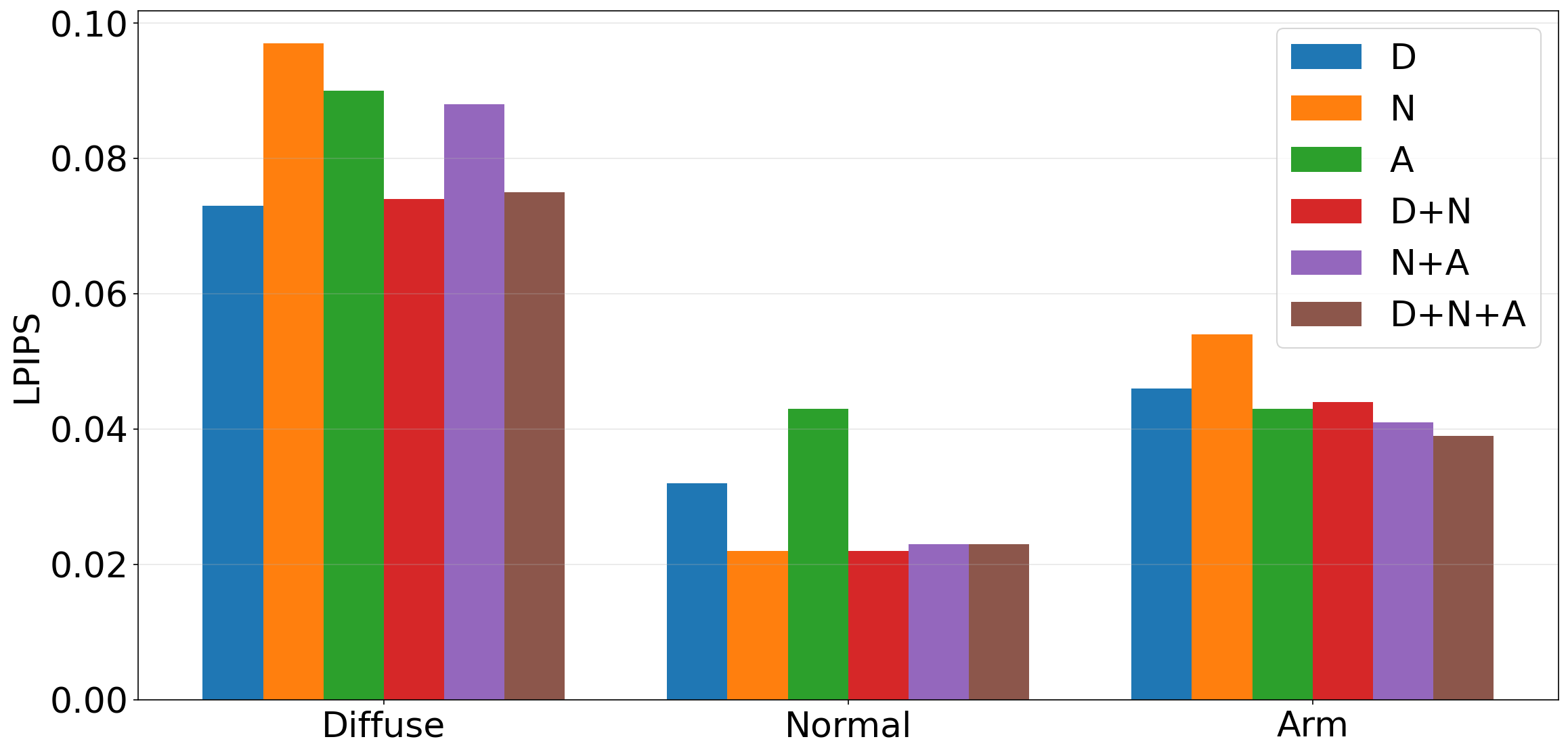}
         \caption{\textsc{Hangar}}
         \label{fig:image_a}
     \end{subfigure}
     \hspace{0.025\textwidth}
     \begin{subfigure}[b]{0.42\textwidth}
         \centering
         \includegraphics[width=\textwidth]{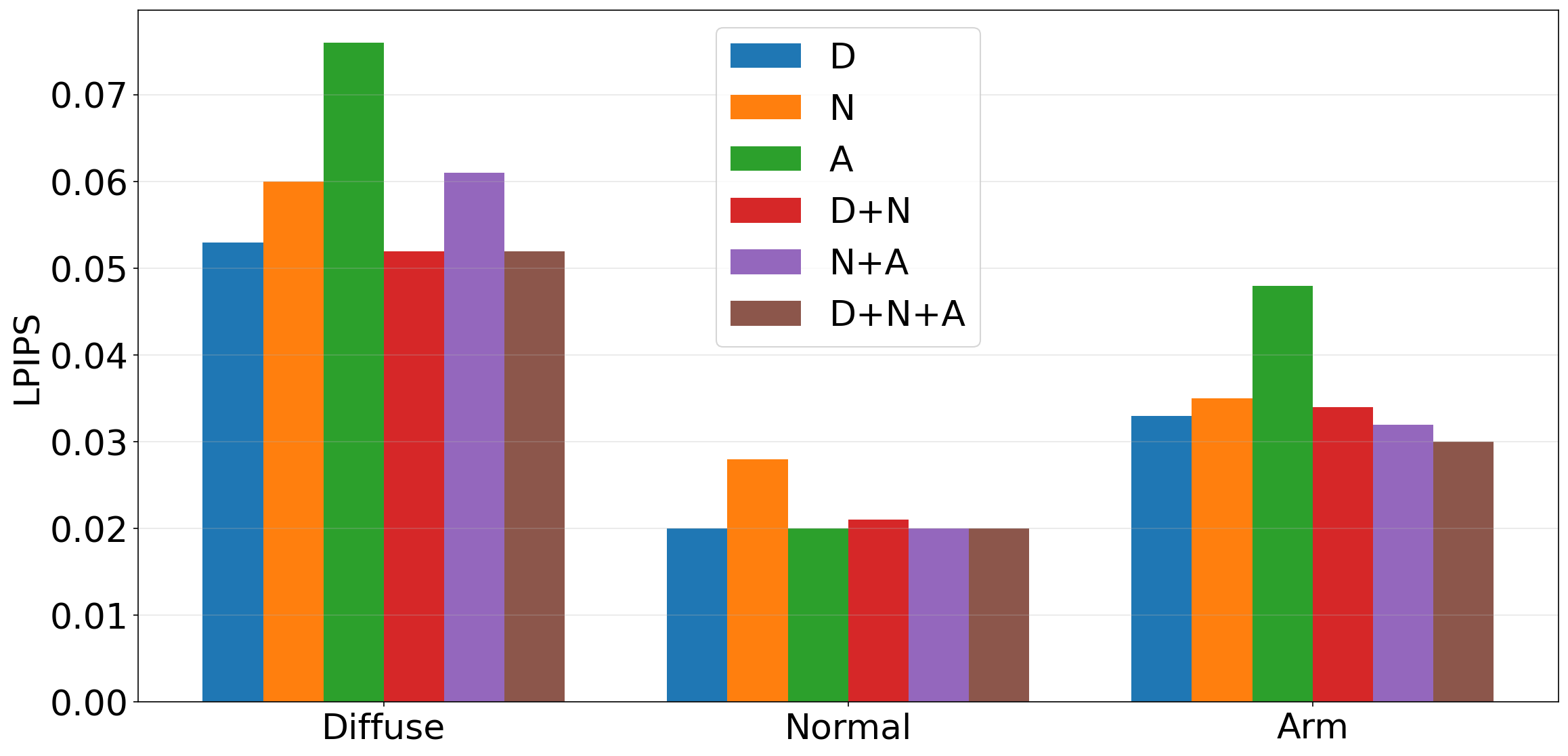}
         \caption{\textsc{Museum}}
         \label{fig:image_b}
     \end{subfigure}     
     \caption{Training dataset study using \textsc{Hangar} and \textsc{Museum} as test data. D, N, A stand for the model trained with diffuse, normal, and ARM textures in the \textsc{Poly Haven} dataset. }
     \label{fig:datastudy}
\end{figure*}

\begin{table*}
  \centering
  \caption{Average metrics for the test datasets using (a) our method, (b) bilinear, (c) bicubic, (d) Lanczos upscaling, and (e) PFT. Rows are categorized by the dataset. }
  \label{tab:per_class_metric}
  \scriptsize
  \begin{tabular}{cc|ccccc|ccccc|ccccc}
    && \multicolumn{5}{c|}{LPIPS$\downarrow$} & \multicolumn{5}{c|}{PSNR$\uparrow$} & \multicolumn{5}{c}{SSIM$\uparrow$} \\
    && (a) & (b) & (c) & (d) & (e) & (a) & (b) & (c) & (d) & (e) & (a) & (b) & (c) & (d) & (e) \\
    \hline
    \hline
    \textsc{Poly Haven} & {Diffuse} & 0.239 & 0.396 & 0.369 & 0.383 & 0.224 & 32.45 & 31.18 & 31.99 & 32.12 & 33.62 & 0.807 & 0.743 & 0.782 & 0.787 & 0.832 \\
    \hline
    \multirow{3}{*}{\textsc{Hangar}} & {Diffuse} & 0.075 & 0.193 & 0.156 & 0.157 & 0.043 & 36.86 & 34.31 & 35.42 & 35.64 & 39.38 & 0.954 & 0.929 & 0.942 & 0.944 & 0.969 \\
    & {Normal} & 0.023 & 0.088 & 0.067 & 0.065 & 0.014 & 37.68 & 35.31 & 36.33 & 36.52 & 39.46 & 0.957 & 0.934 & 0.945 & 0.946 & 0.968 \\
    & {ARM} & 0.041 & 0.094 & 0.068 & 0.065 & 0.024 & 36.64 & 35.30 & 36.76 & 37.05 & 39.23 & 0.922 & 0.924 & 0.938 & 0.940 &  0.952\\
    \hline
    \multirow{3}{*}{\textsc{Victorian}} & {Diffuse} & {0.1026} & 0.2254 & 0.1895 & 0.1918 & 0.091 & {37.37} & 35.32 & 36.42 & 36.60 & 39.12 & {0.9424} & 0.9135 & 0.9311 & 0.9339 & 0.954\\
    & {Normal}  & {0.0246} & 0.0866 & 0.0662 & 0.0651 & 0.021 & 37.65 & 37.56 & 39.00 & {39.23} & 41.71 & 0.9555 & 0.9444 & 0.9563 & {0.9582} & 0.974\\
    & ARM     & {0.0465} & 0.1139 & 0.0825 & 0.0809 & 0.023 & {40.75} & 39.05 & 40.45 & 40.71 & 42.29 & 0.9494 & 0.9295 & 0.9472 & {0.9497} & 0.960\\
    \hline
    \multirow{3}{*}{\textsc{Museum}} & {Diffuse} & 0.052 & 0.142 & 0.107 & 0.107 & 0.048 & 40.43 & 38.30 & 40.03 & 40.41 & 41.95 & 0.968 & 0.947 & 0.963 & 0.965 & 0.975 \\
    & {Normal} & 0.020 & 0.064 & 0.044 & 0.042 & 0.012  & 38.44 & 38.61 & 40.23 & 40.55 & 42.84 & 0.966 & 0.961 & 0.970 & 0.971 & 0.982 \\
    & {ARM} & 0.031 & 0.085 & 0.062 & 0.061 & 0.020 & 38.30 & 37.37 & 39.13 & 39.54 & 41.62 & 0.946 & 0.934 & 0.951 & 0.954 & 0.968 \\
    \hline
  \end{tabular}
\end{table*}

\begin{figure}
     \centering
         \includegraphics[width=\columnwidth]{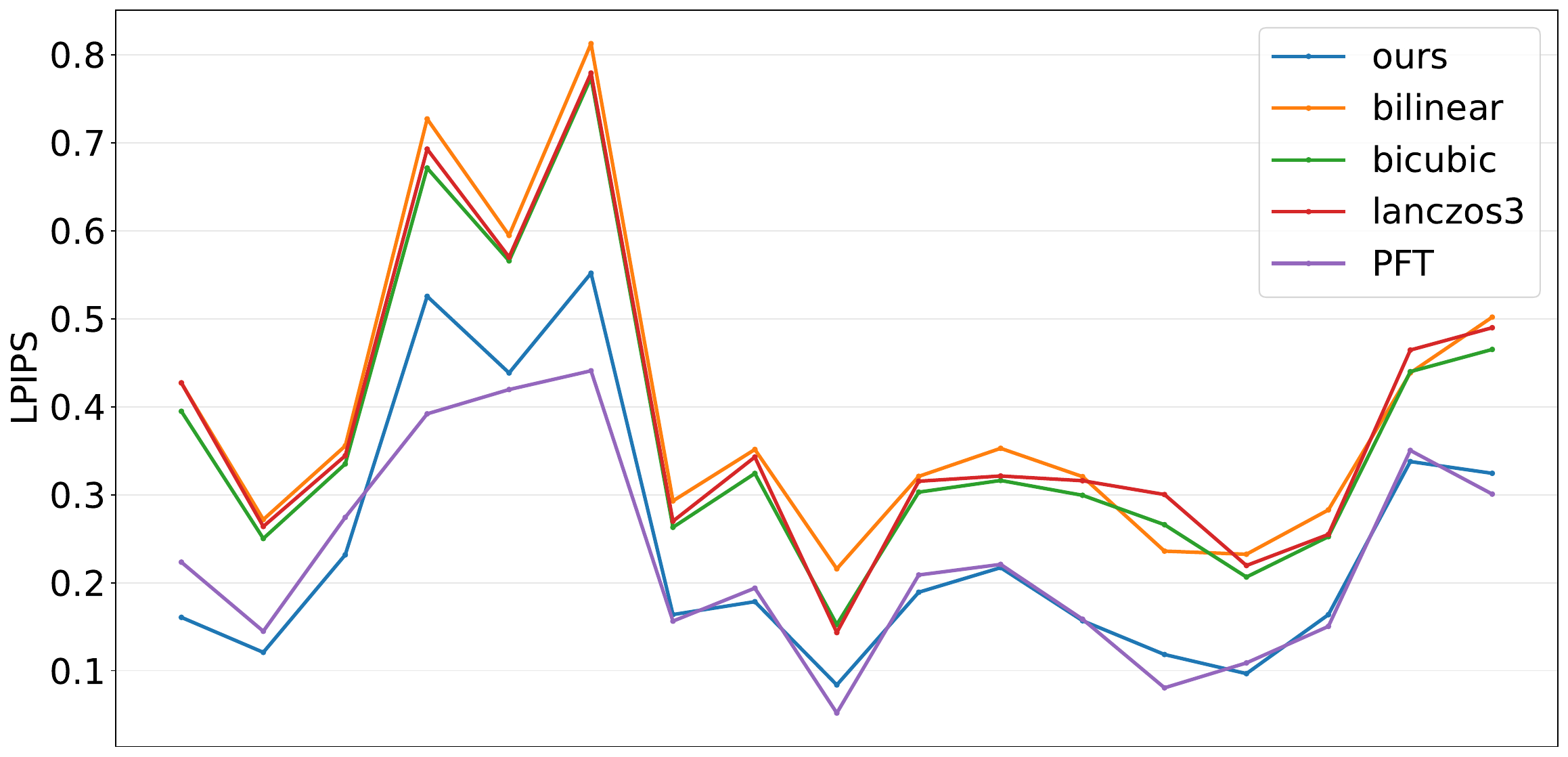}
     \caption{Per-texture LPIPS comparison for textures in the diffuse class from the \textsc{Poly Haven} dataset.}
     \label{fig:ply_diffuse}
\end{figure}

\begin{figure*}
     \centering
     \begin{subfigure}[b]{0.32\textwidth}
         \centering
         \includegraphics[width=\textwidth]{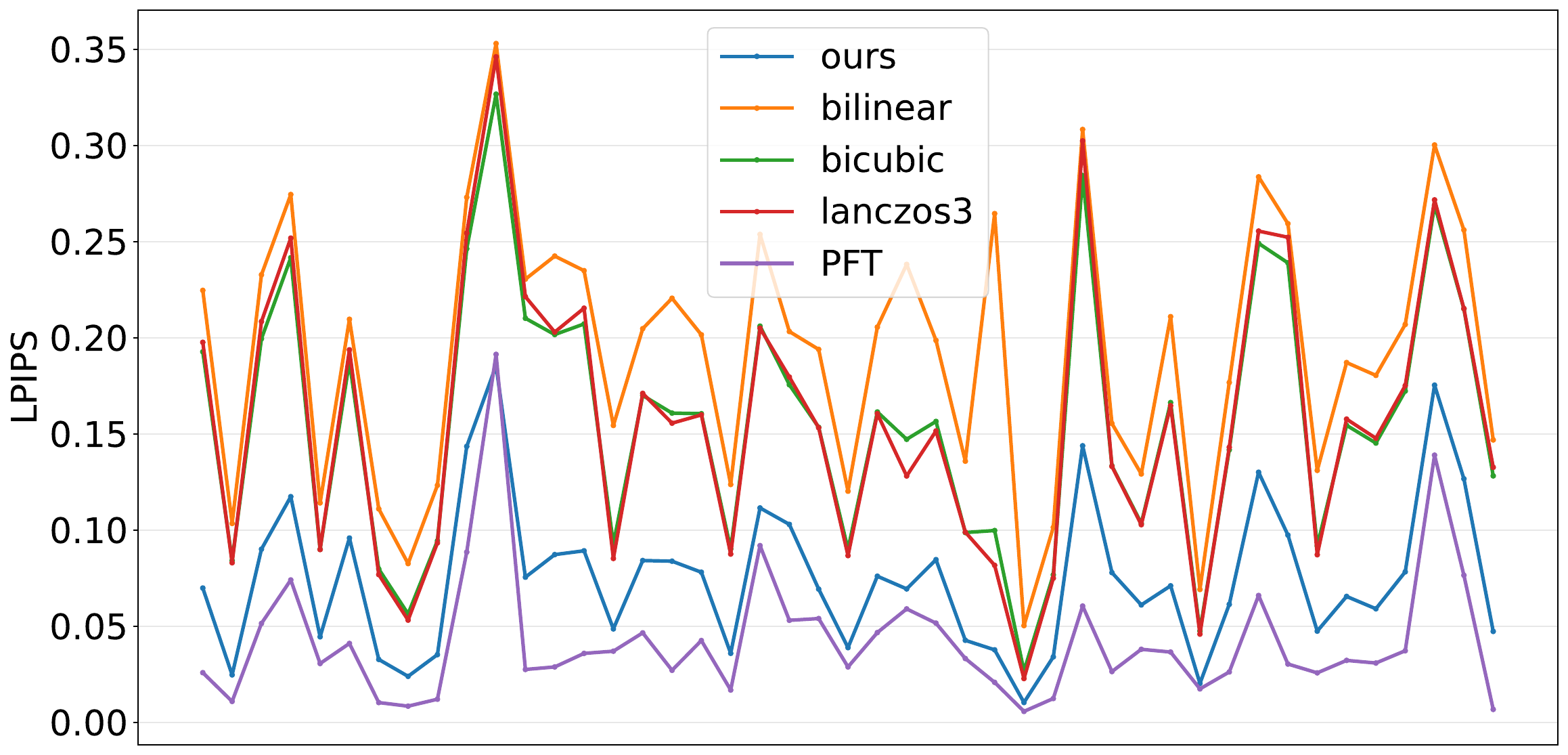}
         \caption{\textsc{Hangar}}
         \label{fig:image_a}
     \end{subfigure}
     \hfill
     \begin{subfigure}[b]{0.32\textwidth}
         \centering
         \includegraphics[width=\textwidth]{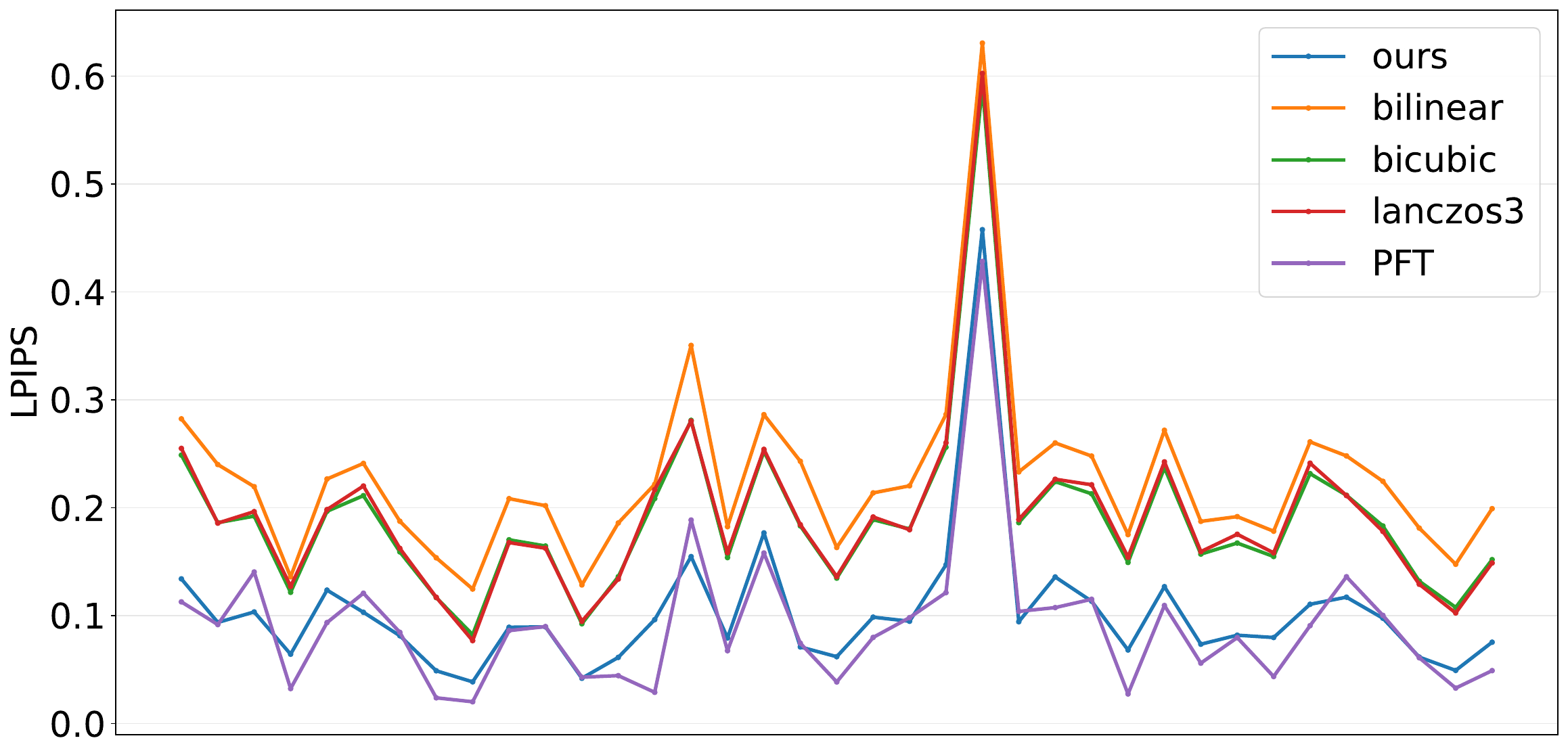}
         \caption{\textsc{Victorian}}
         \label{fig:image_a}
     \end{subfigure}
     \hfill
     \begin{subfigure}[b]{0.32\textwidth}
         \centering
         \includegraphics[width=\textwidth]{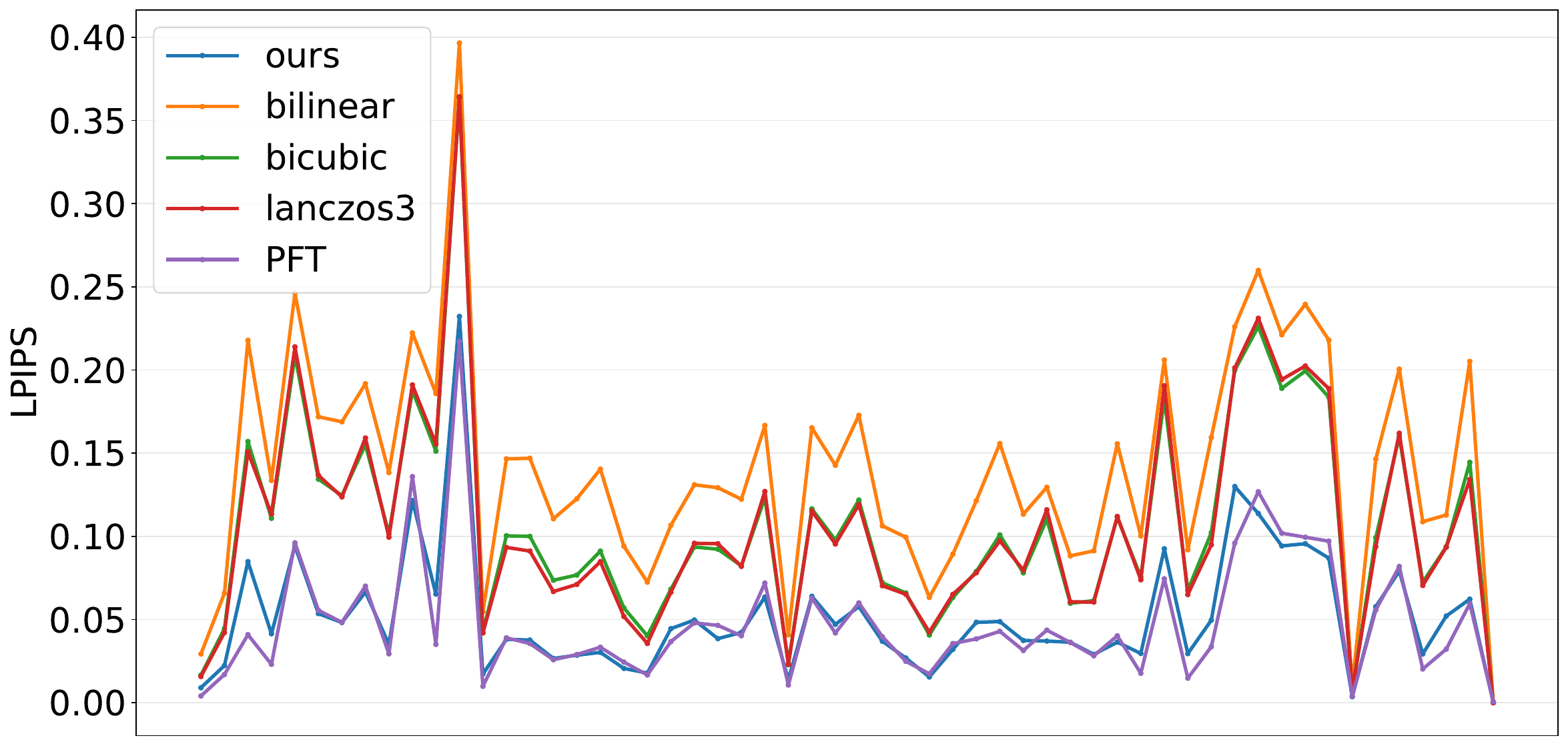}
         \caption{\textsc{Museum}}
         \label{fig:image_b}
     \end{subfigure}     
     \caption{Per-texture LPIPS comparisons for textures in the diffuse class. }
     \label{fig:diffuse}
\end{figure*}

\begin{figure*}
     \centering
     \begin{subfigure}[b]{0.32\textwidth}
         \centering
         \includegraphics[width=\textwidth]{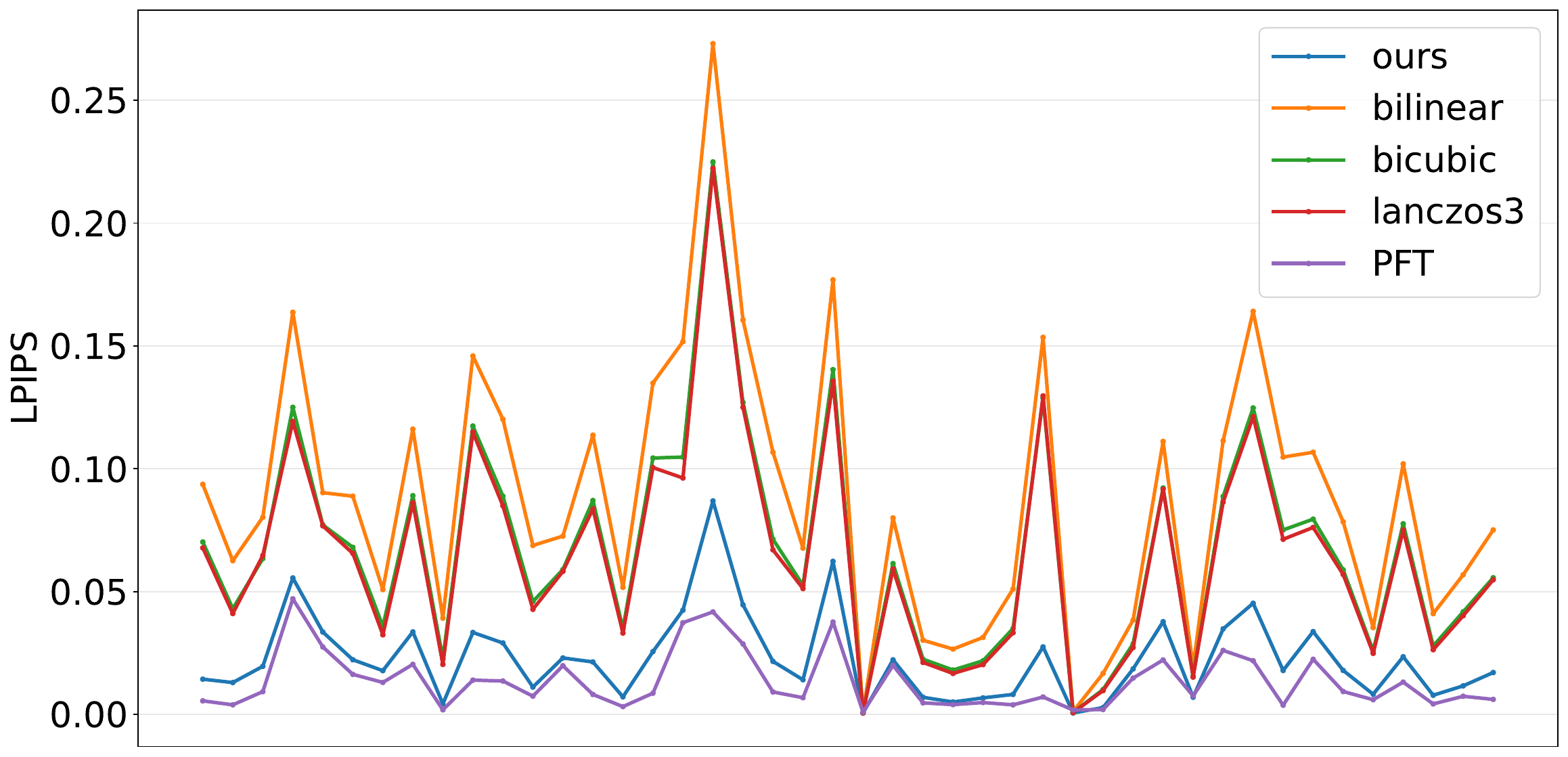}
         \caption{\textsc{Hangar}}
         \label{fig:image_a}
     \end{subfigure}
     \hfill
     \begin{subfigure}[b]{0.32\textwidth}
         \centering
         \includegraphics[width=\textwidth]{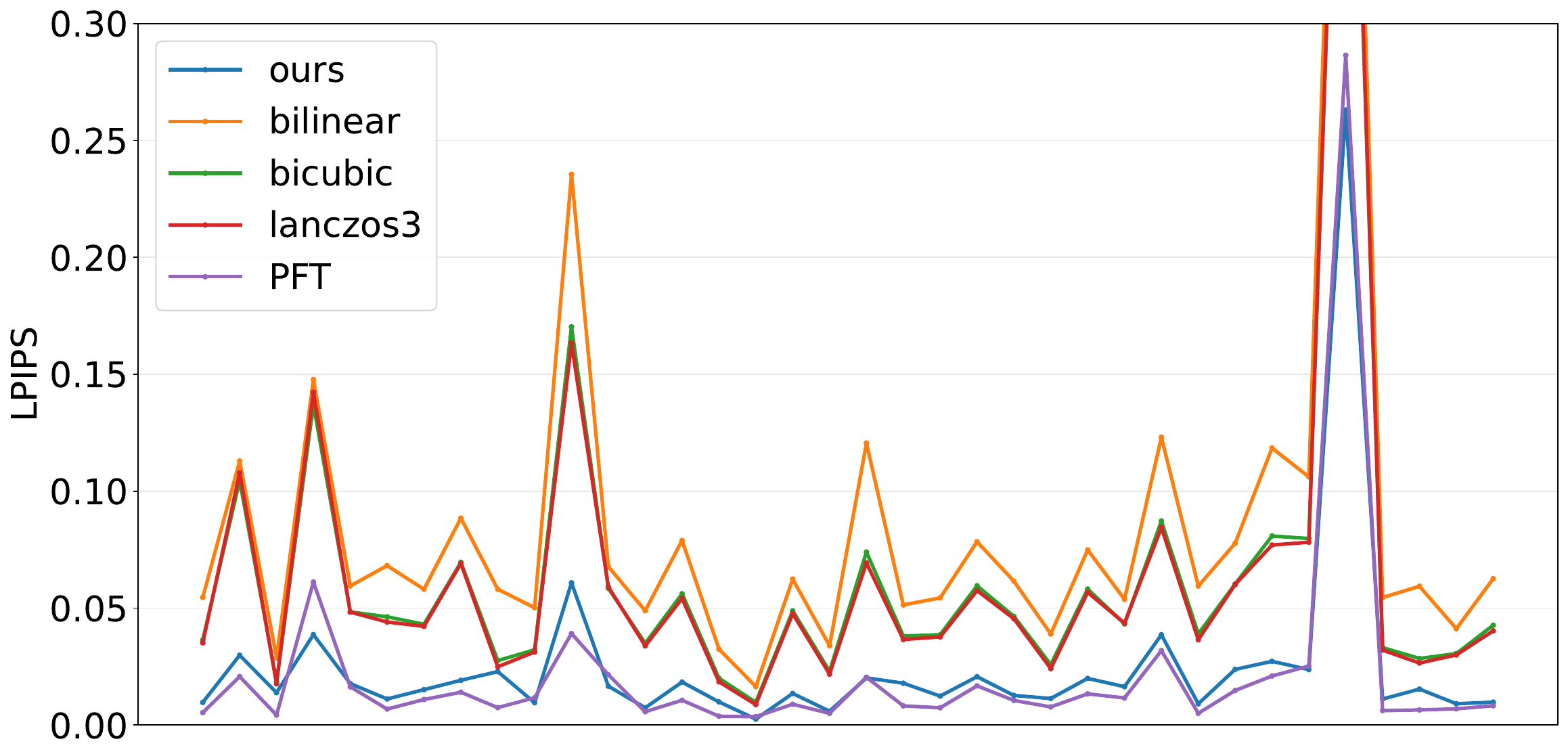}
         \caption{\textsc{Victorian}}
         \label{fig:image_a}
     \end{subfigure}
     \hfill
     \begin{subfigure}[b]{0.32\textwidth}
         \centering
         \includegraphics[width=\textwidth]{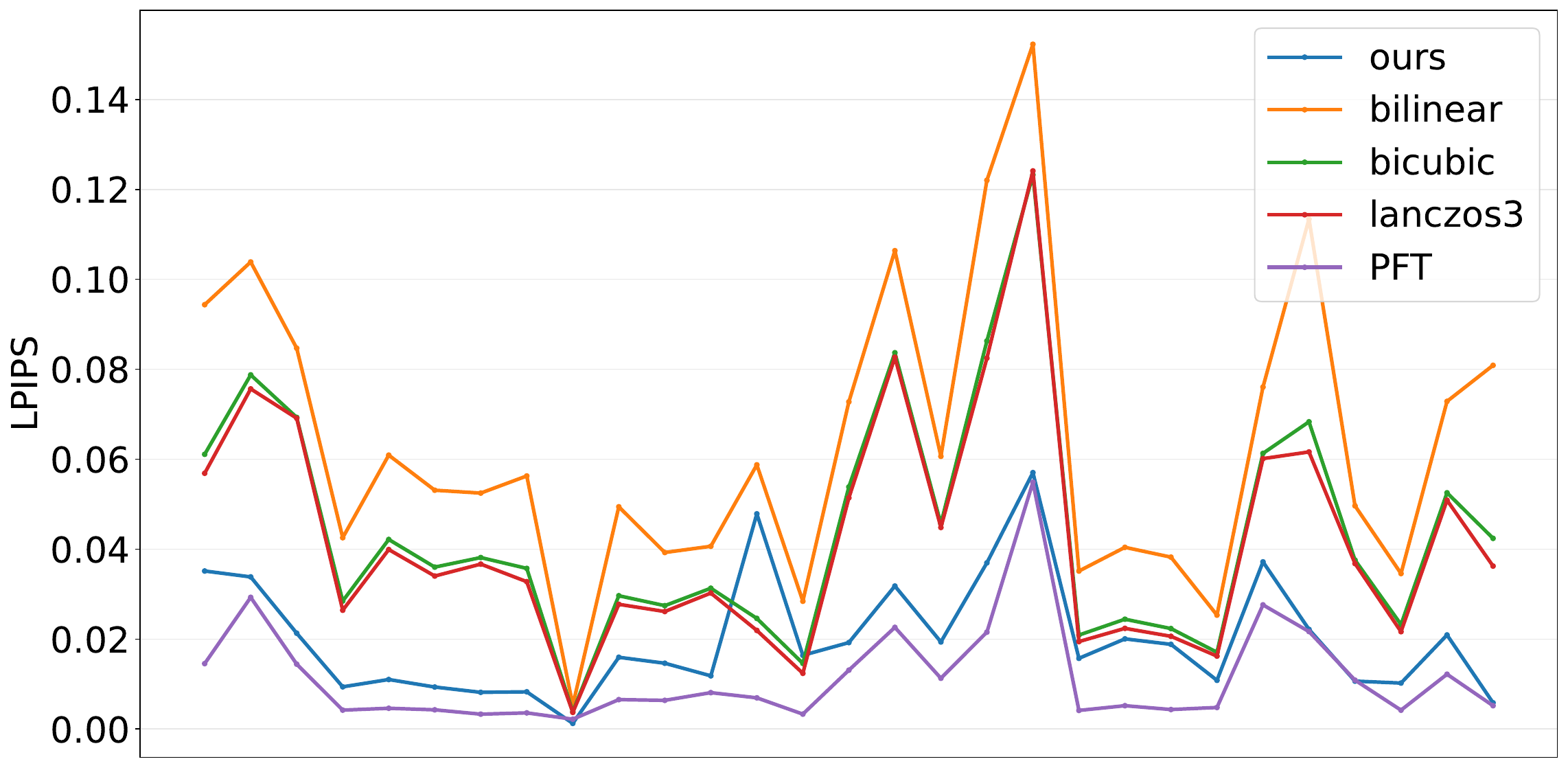}
         \caption{\textsc{Museum}}
         \label{fig:image_b}
     \end{subfigure}     
     \caption{Per-texture LPIPS comparisons for textures in the normal class. }
     \label{fig:normal}
\end{figure*}

\begin{figure*}
     \centering
     \begin{subfigure}[b]{0.32\textwidth}
         \centering
         \includegraphics[width=\textwidth]{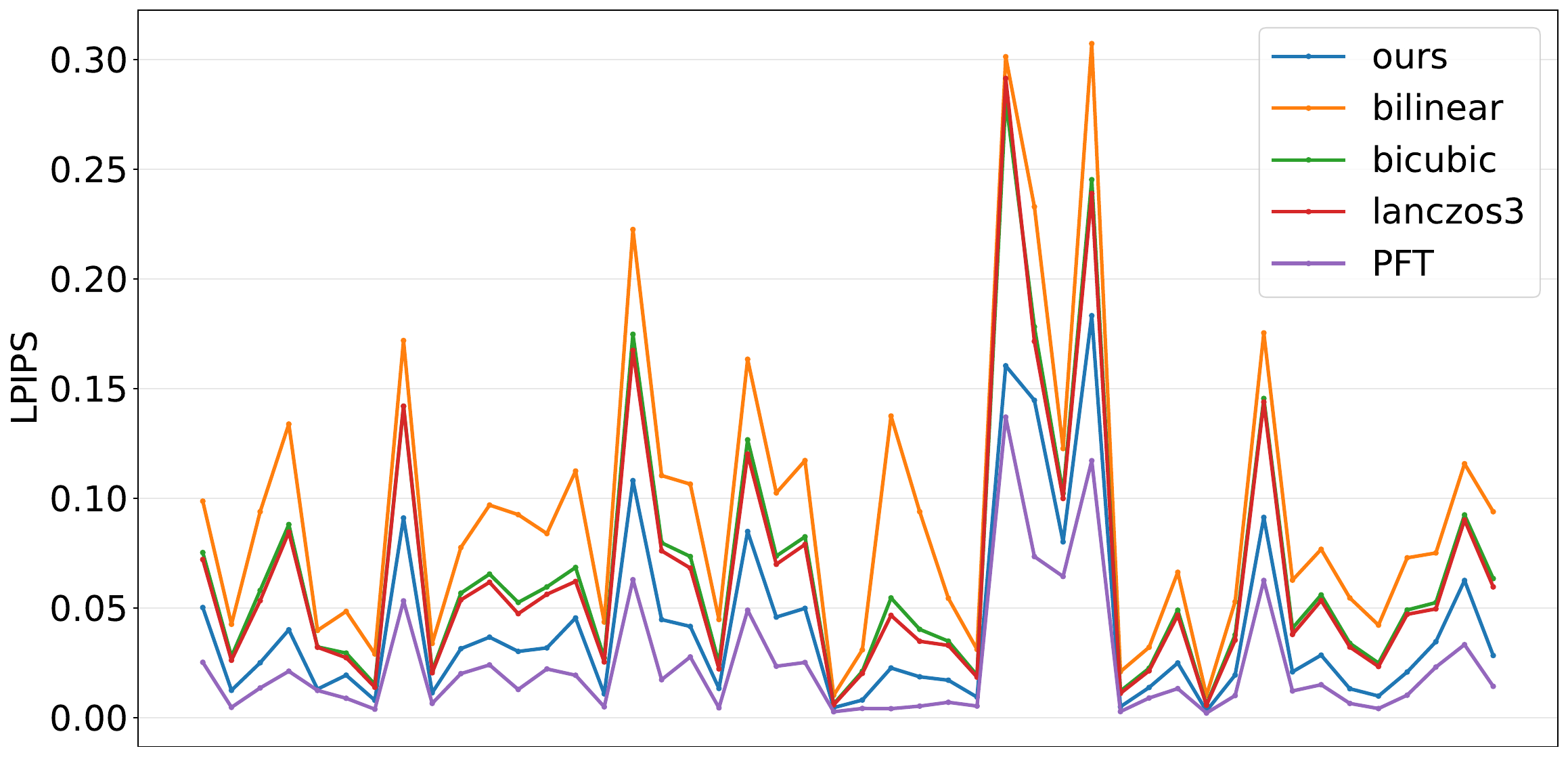}
         \caption{\textsc{Hangar}}
         \label{fig:image_a}
     \end{subfigure}
     \hfill
     \begin{subfigure}[b]{0.32\textwidth}
         \centering
         \includegraphics[width=\textwidth]{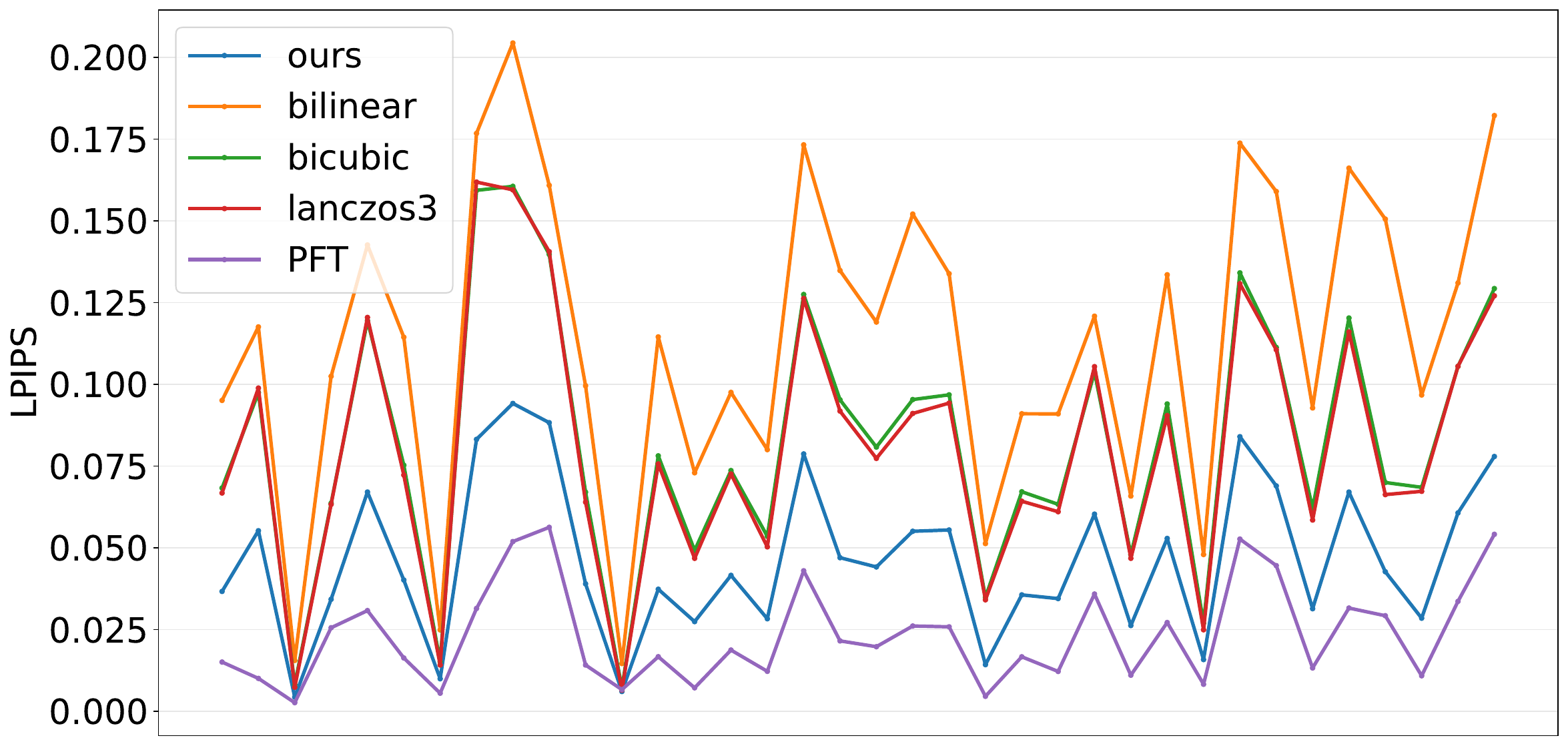}
         \caption{\textsc{Victorian}}
         \label{fig:image_b}
     \end{subfigure}     
     \hfill
     \begin{subfigure}[b]{0.32\textwidth}
         \centering
         \includegraphics[width=\textwidth]{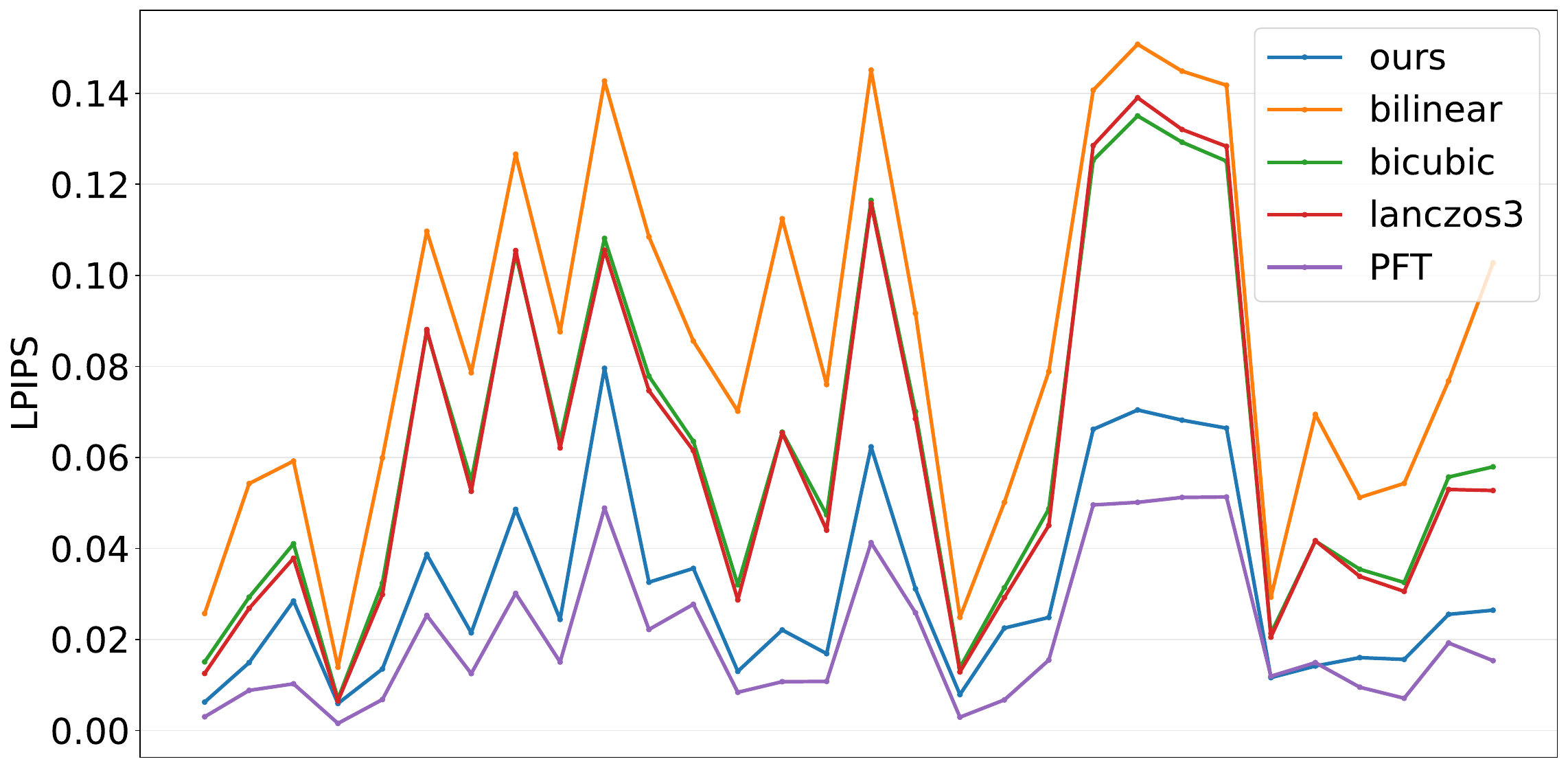}
         \caption{\textsc{Museum}}
         \label{fig:image_b}
     \end{subfigure}     
     \caption{Per-texture LPIPS comparisons for textures in the ARM class.}
     \label{fig:arm}
\end{figure*}

\section{Results}

\subsection{Dataset and Training Configuration}
The model configuration we used for the experiments other than ablation studies is $d=256$ hidden dimension, with residual connections on, $L=4$ residual blocks, $n=1$, and $m=1$, so it produces $2 \times 2$ texel patches. We prepared a few datasets for the experiments as shown in Table~\ref{tab:datasets}. One is taken from the \textsc{Poly Haven} website~\cite{polyhaven2026}, and others are textures from rendering scenes (\textsc{Hangar}, \textsc{Victorian}, \textsc{Museum}). When the \textsc{Poly Haven} dataset is used for testing, the entire dataset is partitioned so that training data does not see the test data. Textures for these datasets are down-sampled to up to $2048\times 2048$ resolution.

We focused on studying upscaling three classes of textures, \textit{diffuse}, \textit{normal}, and \textit{ARM} textures, using the same model architecture. Since these textures show different characteristics, it is natural to train three different models for each dataset. However, if we can unify the models into a single one, it makes its deployment easier than the per-class-model approach. Therefore, we first explored the possibility of preparing the unified models to reduce the total number of models. Fig.~\ref{fig:datastudy} shows the study in which we use \textsc{Hangar} and \textsc{Museum} datasets as test data and compare models trained on different combinations of the \textsc{Poly Haven} training dataset. Here, we used LPIPS (the lower, the better) as the metric in the graphs. For the \textsc{Hangar} dataset, diffuse and normal textures showed the best metric when using the trained model with the same-category training dataset. Diffuse textures in the \textsc{Museum} dataset follow the same pattern, but normal textures did not perform well when the model is trained only with normal textures. In general, using the diffuse + normal + ARM textures for training shows a good balance for both scenes, although they are not the best for each texture class. Therefore, we use a single model trained with the diffuse + normal + ARM textures in the \textsc{Poly Haven} training dataset for all the experiments in this paper.

\subsection{Comparison with Baseline Methods}
\label{sec:comparison}
We evaluated the quality of upscaling with our method by comparing different existing methods. We used three major analytic methods (bilinear, bicubic, Lanczos) and PFT, a state-of-the-art machine learning-based high-quality offline upscaler taken from the authors' implementation~\cite{long2025pftsr, long2025progressive}. Because the complexity of PFT is completely different from our method, we expect PFT to be used as the best-case performance for our model. We measured the metrics for each texture in the test dataset. Table~\ref{tab:per_class_metric} shows average metrics per texture class for each dataset. Although there are no huge differences in PSNR or SSIM, our method and PFT make relatively large differences in LPIPS. As the upscaled textures are not guaranteed to be pixel-accurate, per-pixel distance metrics such as PSNR are not well-suited to evaluate our method.
To prove it, we confirmed that the LPIPS perceptual metric follows naturally with our visual evaluation, rather than per-pixel metrics. We found many cases where blurred textures score higher PSNR values than sharp textures, which we visually find higher quality. This is because PSNR only considers the error in each pixel rather than perceptual correctness. 
Therefore, we use LPIPS as the main metric for further discussion in this paper.

\begin{figure}
     \centering
     \begin{subfigure}[b]{\columnwidth}
         \centering
         \includegraphics[width=\textwidth]{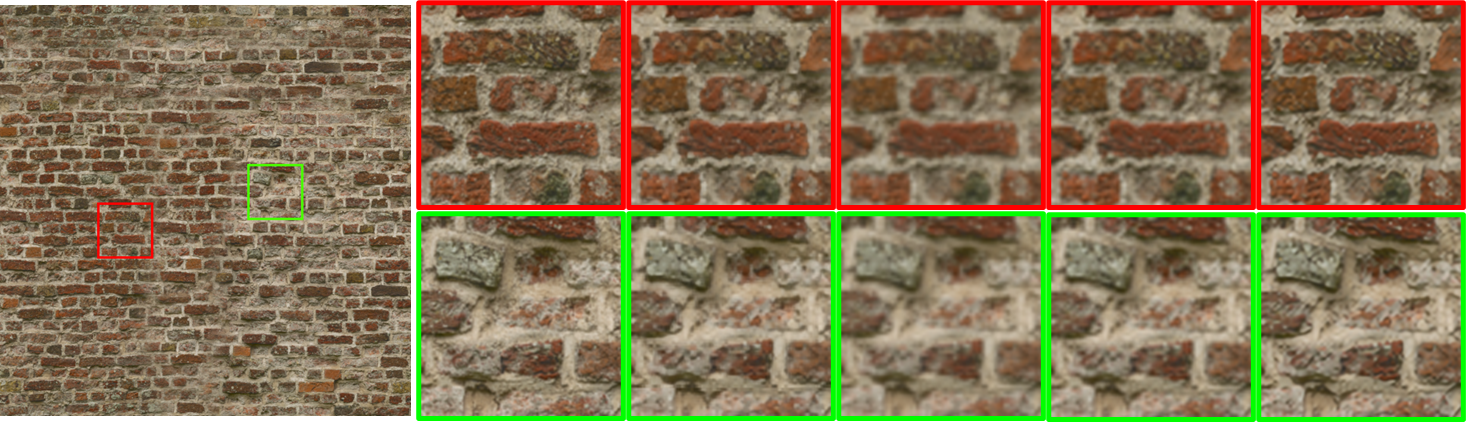}
         \caption{castle\_brick\_02\_white (0.1607, 0.4271, 0.3949, 0.2235).}
         \label{fig:image_a}
     \end{subfigure}
     \hfill
     \begin{subfigure}[b]{\columnwidth}
         \centering
         \includegraphics[width=\textwidth]{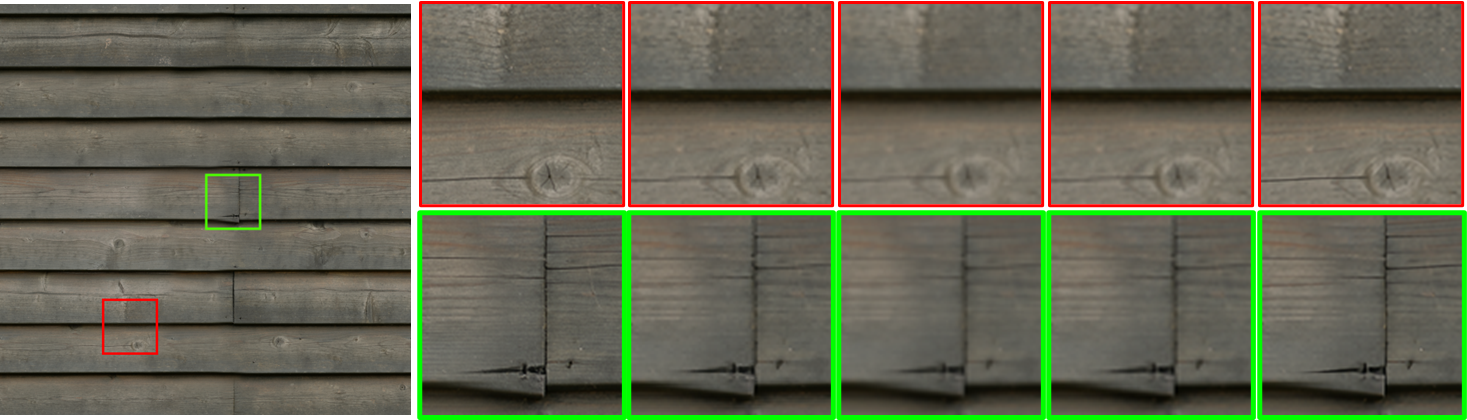}
         \caption{dark\_planks (0.1637, 0.2828, 0.2524, 0.1504).}
         \label{fig:image_b}
     \end{subfigure}
     \begin{subfigure}[b]{\columnwidth}
         \centering
         \includegraphics[width=\textwidth]{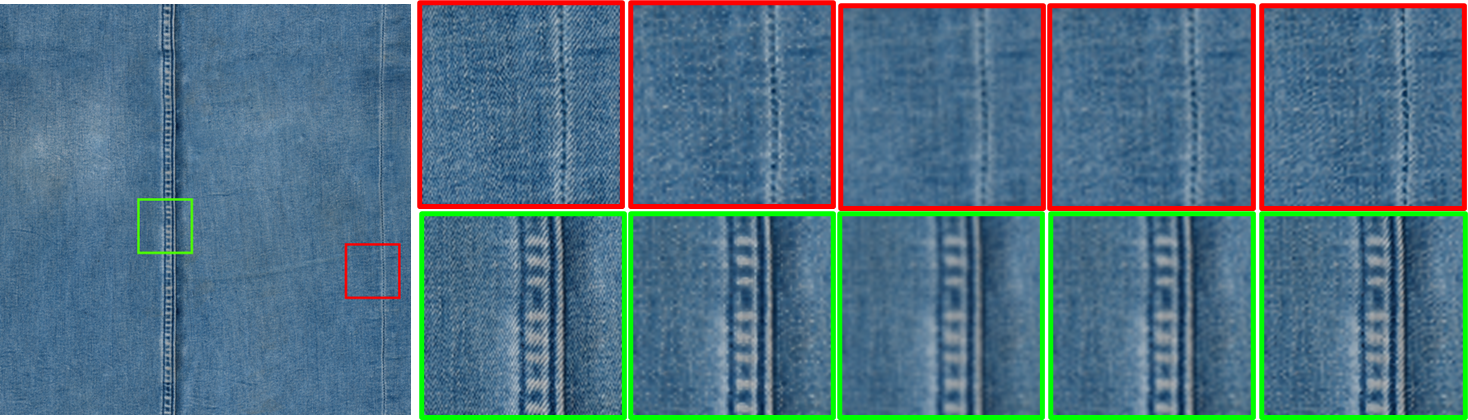}
         \caption{denim\_fabric (0.5518, 0.8128, 0.7742, 0.4409). }
         \label{fig:image_b}
     \end{subfigure}     
     \caption{Visual comparisons of upscaled diffuse textures from \textsc{Poly Haven} dataset. From left to right, reference, our method, bilinear, bicubic, PFT. Numbers in the captions are LPIPS metrics.}
     \label{fig:vis0}
\end{figure}

\begin{figure}
     \centering
     \begin{subfigure}[b]{\columnwidth}
         \centering
         \includegraphics[width=\textwidth]{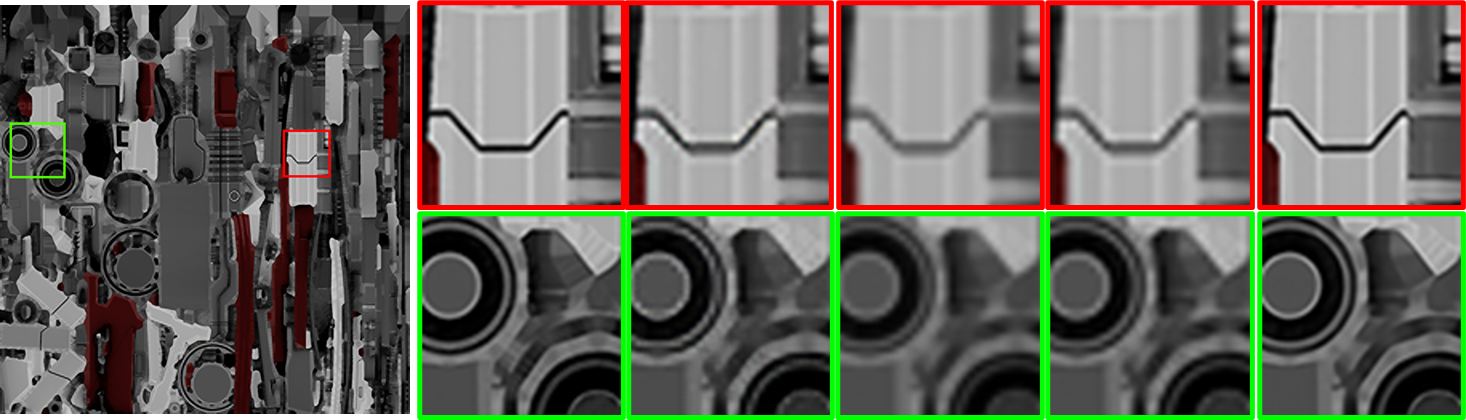}
         \caption{Weapon\_D (0.0974, 0.2594, 0.2390, 0.0304).}
         \label{fig:image_a}
     \end{subfigure}
     \hfill
     \begin{subfigure}[b]{\columnwidth}
         \centering
         \includegraphics[width=\textwidth]{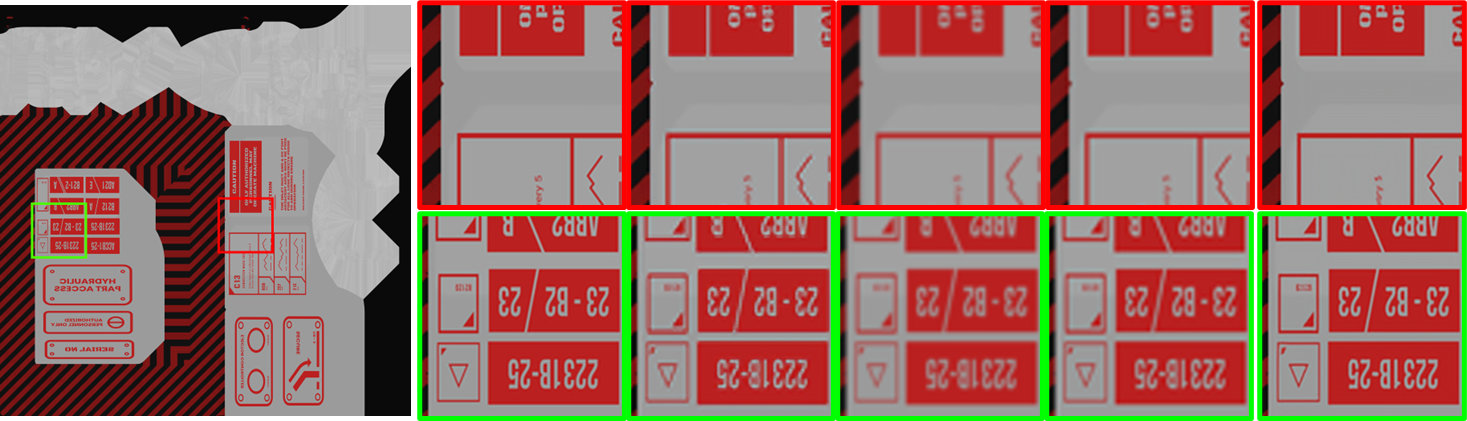}
         \caption{Hangar\_Decals\_BaseColor (0.0359, 0.1238, 0.0902, 0.0168).}
         \label{fig:image_b}
     \end{subfigure}
     \caption{Visual comparisons of upscaled diffuse textures from \textsc{Hangar} dataset. From left to right, reference, our method, bilinear, bicubic, PFT. Numbers in the captions are LPIPS metrics.}
     \label{fig:vis1}
\end{figure}

\begin{figure}
     \centering
     \begin{subfigure}[b]{\columnwidth}
         \centering
         \includegraphics[width=\textwidth]{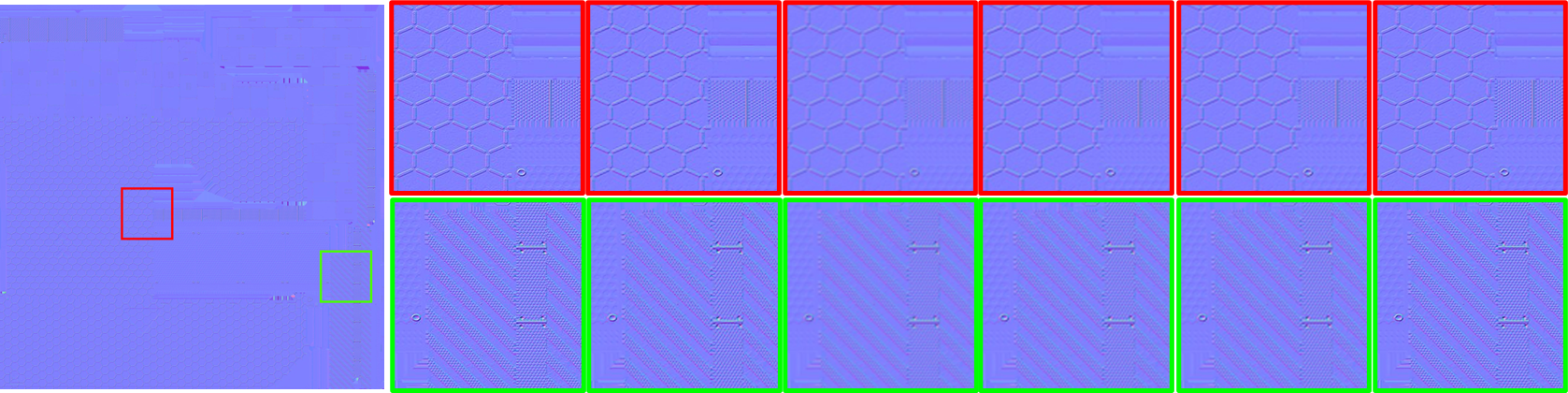}
         \caption*{Hangar-FloorTile-v1\_Normal (0.0869, 0.2731, 0.2249, 0.2223, 0.0417).}
         \label{fig:image_a}
     \end{subfigure}
     \caption{Visual comparison of upscaled normal textures from \textsc{Hangar} dataset. From left to right, reference, our method, bilinear, bicubic, Lanczos, PFT. Numbers in the captions are LPIPS metrics. }
     \label{fig:vis3}
\end{figure}

\begin{table}[t]
  \centering
  \caption{FLIP errors for rendered images shown in Fig.~\ref{fig:teaser} and Fig.~\ref{fig:hangar_cam} using different upscaling methods. }
  \label{tab:flip}
  \small
  \begin{tabular}{lccc}
  \toprule
  Scene & Ours & Bicubic & Bilinear \\
  \midrule
  \textsc{Hanger} (Camera 1) & \textbf{0.0519} & 0.0573 & 0.0626 \\
  \textsc{Hangar} (Camera 2) & \textbf{0.0415} & 0.0441 & 0.0473 \\
  \textsc{Hanger} (Camera 3) & \textbf{0.0441} & 0.0470 & 0.0495 \\
  \bottomrule
  \end{tabular}
\end{table}

\begin{figure}
     \centering
     \begin{subfigure}[b]{0.495\columnwidth}
         \centering
         \includegraphics[width=\textwidth]{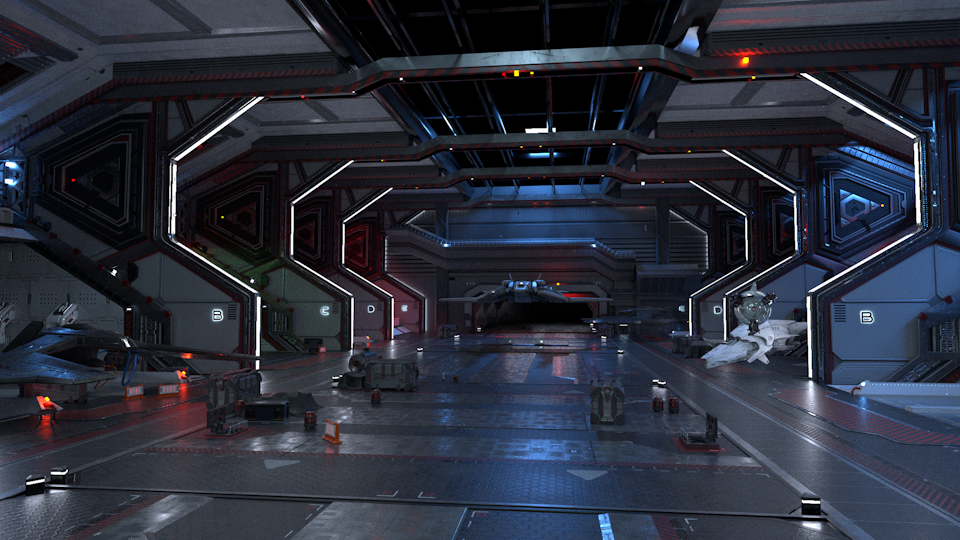}
         \label{fig:image_a}
     \end{subfigure}
     \hfill
     \begin{subfigure}[b]{0.495\columnwidth}
         \centering
         \includegraphics[width=\textwidth]{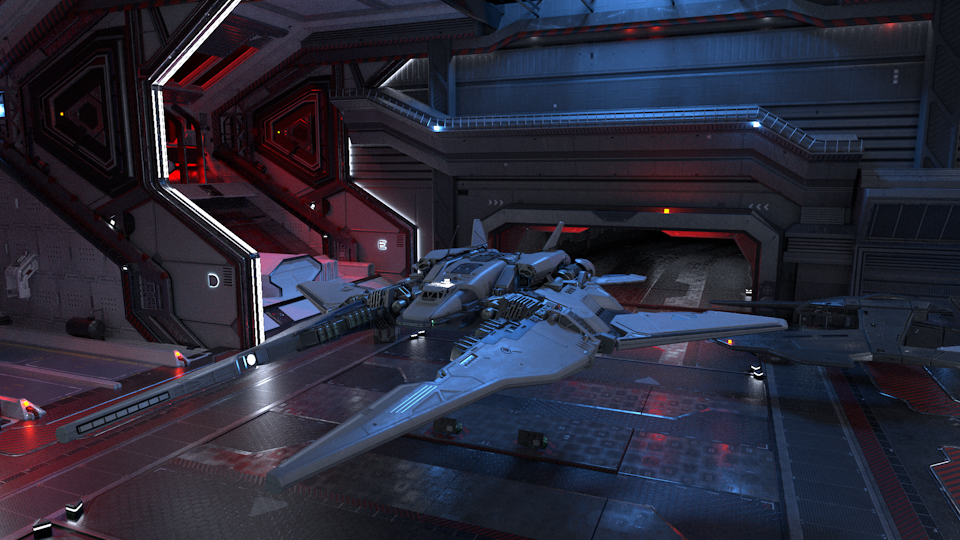}
         \label{fig:image_b}
     \end{subfigure}
     \caption{\textsc{Hangar} scene rendered using Camera 2 and 3. }
     \label{fig:hangar_cam}
\end{figure}

Along with the per-class average metrics, we also measured per-texture metrics for the test datasets used in Table~\ref{tab:per_class_metric}. Fig.~\ref{fig:ply_diffuse},~\ref{fig:diffuse},~\ref{fig:normal}, and~\ref{fig:arm} show the LPIPS metrics for upscaled textures using the proposed method, analytic upscaling methods (bilinear, bicubic, Lanczos), and PFT. We can see that our method achieves better metrics compared to most of the textures upscaled using analytic methods. Our method even matched PFT for some textures from the \textsc{Hangar} normal, \textsc{Victorian} normal, and \textsc{Museum} diffuse datasets, even though the model complexity is totally different. Visual comparison also matches the LPIPS metrics, as illustrated in Fig.~\ref{fig:vis0},~\ref{fig:vis1}, and~\ref{fig:vis3}. All the analytic methods produce smoother results, while our method and PFT produce sharper images, which is especially obvious from Fig.~\ref{fig:vis1}. The same pattern is observed in normal textures, which often have sharp edges and detailed features as shown in Fig.~\ref{fig:vis3}. Lanczos upscaling method, known for sharper reconstruction, also does not make a huge difference to bicubic in the experiments. However, our method reconstructs sharper edges as PFT. 
At the same time, careful inspection of these visual results clarifies the limitation of our method. It is obvious, especially from Fig.~\ref{fig:vis0} (c) and Fig.~\ref{fig:vis3}, that fine detail, less than pixel scale, which does not exist in the source textures, does not appear in any of the upscaled images, even with PFT. This is a limitation of upscaling methods like ours, which do not store any other information than a low-resolution texture. 
We also evaluated our method by rendering scenes with a different set of textures using our renderer written in HIP~\cite{hip}. One example is shown in Fig.~\ref{fig:teaser}, where our method produces fewer errors, resulting in a low FLIP score as shown in Table~\ref{tab:flip}. The difference is obvious in the glossy reflection. We can even see the difference coming from diffuse textures in Fig.~\ref{fig:teaser}. The numbers in Table~\ref{tab:flip} are from different cameras, and the teaser is rendered using the first camera. Other views are shown in Fig.~\ref{fig:hangar_cam}. The difference is bigger for views where the camera is close to objects, where we can see textures clearly. 

Although we have not explored it in this research, we could omit the MLP execution completely and just sample the low-resolution texture at the vertices where high quality is not required, e.g., vertices after a diffuse reflection. This is possible with our method, as the input to the NN is the texels, while NTC cannot do so because the latent grid needs to be decoded before use.

\begin{table}
  \centering
  \caption{LPIPS for overfit experiments.}
  \small
  \label{tab:w24_overfit_lpips}
  \begin{tabular}{lrr}
  \toprule
  train and test datasets & D+N+A & overfit \\
  \midrule
  Hangar (D)     & 0.076 & 0.059 \\
  Hangar (N)     & 0.023 & 0.020 \\
  Hangar (A)     & 0.041 & 0.031 \\
  Victorian (D)  & 0.103 & 0.091 \\
  Victorian (N)  & 0.025 & 0.023 \\
  Victorian (A)  & 0.047 & 0.033 \\
  Museum (D)     & 0.052 & 0.047 \\
  Museum (N)     & 0.020 & 0.021 \\
  Museum (A)     & 0.032 & 0.026 \\
  \bottomrule
  \end{tabular}
\end{table}

\subsection{Generalization vs. Overfitting}
The results we have seen are obtained using generalized models: a single model is trained and applied to different test textures. Under some situations, we may want to overfit a model to the texture set as NTC does. It requires more training, but it should improve the quality. There are many options, but a reasonable option is to train a model for each class using the textures it needs to upscale. We compared the results with an over-fitted model to the ones with a generalized model in Table~\ref{tab:w24_overfit_lpips}. 
Although the improvement with an over-fitted model is consistent except in a few cases, such as \textsc{Museum} normal, where the generalized model performed better, over-fitting adds overhead for the content creation pipeline. Whether to use overfitting or a generalized model depends on the situation of the development. Thus, developers can choose the appropriate approach case by case.

\begin{table*}
    \centering
    \caption{LPIPS comparison across hidden dimensions and analytic methods. }
    \small
    \label{tab:hidden_dim}
    \begin{tabular}{lrrrrrrrrr}
    \toprule
     & \multicolumn{7}{c}{Hidden dim} & \multicolumn{2}{c}{Analytic} \\
    \cmidrule(lr){2-8} \cmidrule(lr){9-10}
     & 512 & 256 & 128 & 64 & 32 & 16 & 8 & Bilinear & Bicubic \\
    \midrule 
      Hangar (D) & 0.075 & 0.075 & 0.078 & 0.083 & 0.090 & 0.086 & 0.079 & 0.193 & 0.156 \\
      Hangar (N) & 0.023 & 0.023 & 0.025 & 0.033 & 0.039 & 0.040 & 0.040 & 0.088 & 0.067 \\
      Hangar (A) & 0.039 & 0.040 & 0.041 & 0.045 & 0.042 & 0.043 & 0.043 & 0.094 & 0.068 \\
    \bottomrule
    \end{tabular}
\end{table*}

\subsection{Comparison to Neural Texture Compression}

\begin{figure}
    \centering
    \includegraphics[width=\columnwidth]{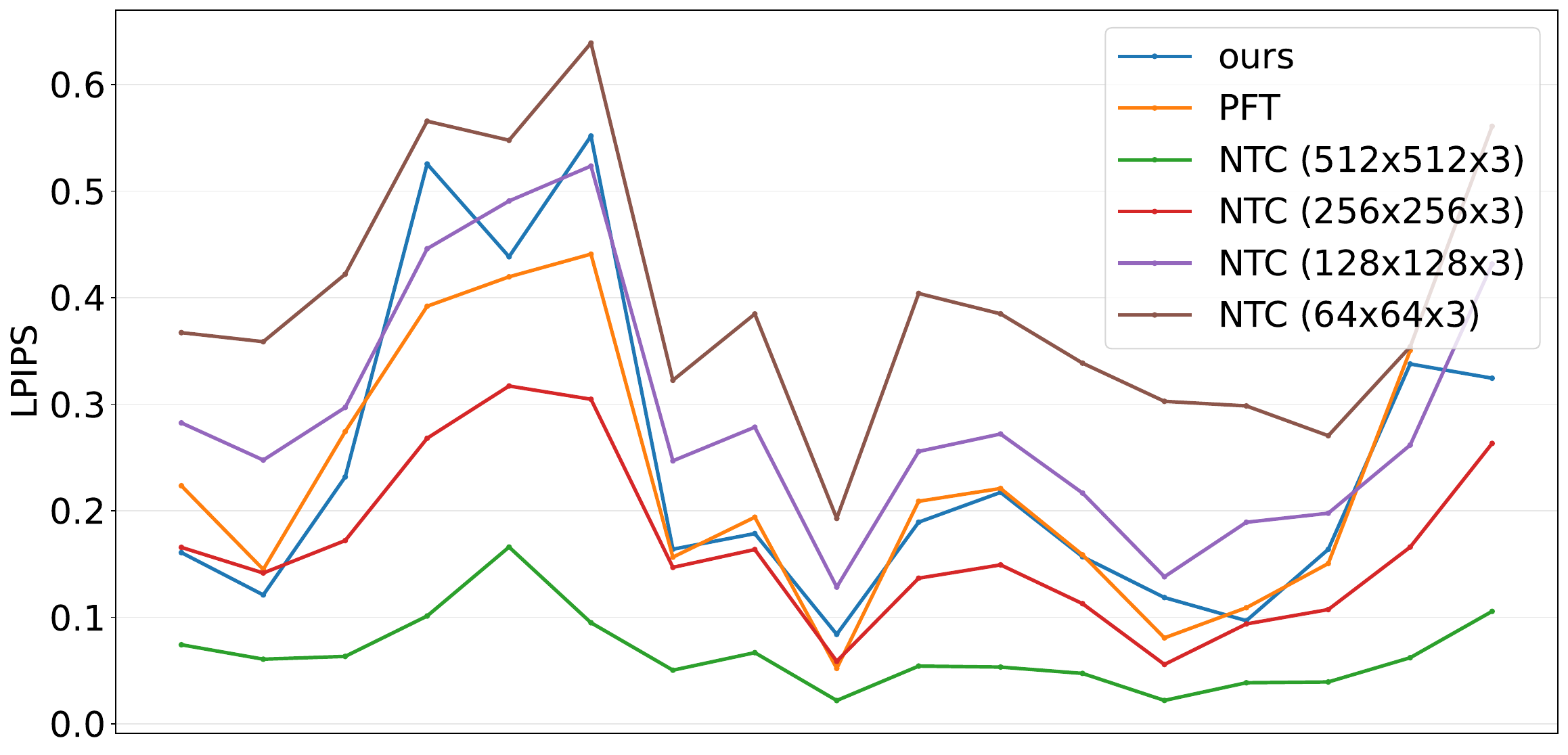}
    \caption{Comparison to NTC variants.}
    \label{fig:ntc}
\end{figure}

We compared our method with NTC, another neural method for texture compression, although our method does not compress the original textures. For NTC, we compressed a single diffuse texture from the \textsc{Poly Haven} test set using Positional Encoding Projected Sampling (PEPS) with four frequencies, which produces $11$-dimensional inputs for an MLP with $64$ hidden dimensions and $4$ hidden layers~\cite{10.1145/3806062}. The comparison is shown in Fig.~\ref{fig:ntc}. The result from NTC ($512 \times 512 \times 3$) is the equal-size comparison using a $512 \times 512 \times 3$ latent grid. Strictly, it is larger, as NTC needs to store an MLP per texture, while our method stores a single MLP for all the textures. However, we ignored it for simplicity as it is relatively small. 
Neither our method nor PFT can match NTC ($512 \times 512 \times 3$). This is mainly because NTC is fitted directly to the ground-truth texture during compression, while our method and PFT do not. Our method roughly matches NTC ($256 \times 256 \times 3$) but falls behind on difficult cases where the low-resolution input likely lacks the details needed for faithful reconstruction. Modifying our network so that it can also take the high-resolution input is likely to improve our reconstruction, but we do not explore this in this work as it changes the assumption that we do not overfit to each texture. 

A different philosophy of the model changes the behavior at runtime. Consider a case where we need to execute a neural network to get texels in the screen space; NTC suffers from divergence because it creates a model for each texture set. Sorting would be required to execute efficiently on GPUs, which complicates and adds overhead to the execution~\cite{10.2312:hpg.20251173}. On the other hand, our method uses a single model for all the textures; divergence does not exist. If we compare our model with NTC having a similar configuration ($64$ hidden dims in Table~\ref{tab:hidden_dim}), our method has the possibility to run faster because there is no divergence. Performance evaluation is future work.

\begin{table}
  \centering
  \caption{Ablation study on hyperparameters tested using diffuse textures from \textsc{Hangar} dataset. The configuration used for other experiments is highlighted in bold under the \textit{Config} column. The best numbers in each experiment are highlighted in bold. }
  \small
  \label{tab:ablation}
  \begin{tabular}{llccc}
  \toprule
  Experiment & Config & LPIPS $\downarrow$ & PSNR $\uparrow$ & SSIM $\uparrow$ \\
  \midrule
  \multirow{2}{*}{Residual connection} & Off & 0.0796 & 36.46 & 0.9494 \\
   & \textbf{On} & \textbf{0.0755} & \textbf{36.87} & \textbf{0.9540} \\
  \midrule
  \multirow{2}{*}{Residual block} & Off & 0.0767 & 36.81 & 0.9535 \\
   & \textbf{On} & \textbf{0.0755} & \textbf{36.87} & \textbf{0.9540} \\
  \midrule
  \multirow{3}{*}{Patch size} & \textbf{1} & \textbf{0.0755} & 36.87 & 0.9540 \\
   & 2 & 0.0767 & \textbf{37.02} & \textbf{0.9555} \\
   & 4 & 0.0899 & 36.57 & 0.9511 \\
  \midrule
  \multirow{4}{*}{Hidden layers} & 1 & 0.0789 & 36.69 & 0.9523 \\
   & 2 & 0.0764 & 36.84 & 0.9536 \\
   & 3 & 0.0761 & 36.86 & 0.9538 \\
   & \textbf{4} & \textbf{0.0755} & \textbf{36.87} & \textbf{0.9540} \\
  \midrule
  \multirow{3}{*}{Padding size} & 0 & 0.0871 & 33.79 & 0.9303 \\
   & \textbf{1} & \textbf{0.0755} & \textbf{36.87} & \textbf{0.9540} \\
   & 2 & 0.0791 & 37.21 & 0.9568 \\
  \bottomrule
  \end{tabular}
\end{table}

\subsection{Ablation Study}

Among the model hyperparameters, the one that affects the texture quality the most is the hidden dimension of the MLP. Table~\ref{tab:hidden_dim} compares the per-class metrics evaluated for the \textsc{Hanger} test data. Obviously, the biggest model with $512$ hidden dimensions performed the best, but the quality degradation is not steep even if we reduce the hidden dimension. If we reduce it to smaller than $128$, the quality decreases faster. This indicates $128$ dimensions is the sweet spot for this model. We can also see that even if we reduce the hidden dimension to $8$, which makes the neural network a tiny MLP that is computationally inexpensive, it still outperforms compared to analytic upscaling methods. 
We also studied the effect of other parameters, which are summarized in Table~\ref{tab:ablation}. The default parameter used in other experiments performed the best among other configurations, except for patch size. Although patch size $4$ performed better than $2$, it requires a larger MLP to keep the quality, which is understandable as the input and output dimensions increase by $4$. The quality was similar when we used a $512$ hidden dimension, but the quality drops fast if we reduce the hidden dimension. A larger patch size would result in the generation of more wasted texels; we used a patch size of $2$ to keep it more practical. Without padding, the quality drops a lot. This is because it makes the input to the neural network a single texel and tries to produce $4$ texels, which is an ill-conditioned problem. We also tried bicubic interpolation for the residual connection, but it did not improve the quality. 

\subsection{Training Data and Model Capacity}
\label{sec:data-and-capacity}

The configuration evaluated so far fixes two choices that the generalized formulation
leaves open: the model is trained on PBR textures, and it uses a $256$-wide network with
four residual blocks. This section is an ablation study of these \footnote{This subsection was added as a response to \url{https://github.com/richgel999/neural_upsampler} (accessed on Nov.4, 2026).}. We first replace the PBR training dataset with natural photographs
(Sec.~\ref{sec:natural-images}), which tests whether the model learns texture statistics; we then reduce the network by two orders of magnitude to see how much quality we need to be sacrificed by this
(Sec.~\ref{sec:capacity}) . 

All comparisons hold the \emph{pixel} budget of the training dataset fixed rather than the
image count, since the number of training patches scales with the number of texels at a
fixed stride. Unless stated otherwise every model is trained for
$250 \times 100 \times 1024 \approx 25.6$M patch draws, so optimization effort is constant
across conditions. Following Sec.~\ref{sec:comparison}, we use LPIPS as the primary metric.

\subsubsection{Training on Natural Images}
\label{sec:natural-images}

PBR datasets are small, licence-encumbered and costly to curate, whereas natural-image
datasets are large and freely available. If the model is learning a generic local
upsampling operator rather than the statistics of materials, photographs should train it
just as well.

We therefore replace the \textsc{Poly Haven} training dataset with \mbox{LSDIR}
\cite{li2023lsdir}, a dataset of 84{,}991 photographs and rendered images collected for
image restoration, and sample from it a subset with the same number of texels as the
\textsc{Poly Haven} diffuse $+$ normal $+$ ARM dataset (1693 maps, $1.779$~Gpixel;
1668 LSDIR images). Nothing else changes. To avoid testing on training data we also build
a held-out \textsc{Poly Haven} split spanning all three map types (17 diffuse, 17 normal,
16 ARM) and retrain the \textsc{Poly Haven} model on the remaining 1646 maps, with an
LSDIR subset matched to that reduced budget.

\begin{table}[t]
\caption{Training on natural images versus PBR textures, per map type, at an equal
training-pixel budget and the same $3\times3$ input window. $\Delta$ is LSDIR minus
\textsc{Poly Haven}; for LPIPS a negative $\Delta$ favours LSDIR. Photographs train
diffuse (D) and ARM (A) maps better on almost every scene and normal maps (N) worse on all four.}
\label{tab:natural-images}
\small
\setlength{\tabcolsep}{4pt}
\begin{tabular}{llrrr@{\hskip 10pt}rrr}
\toprule
 & & \multicolumn{3}{c}{PSNR\,$\uparrow$} & \multicolumn{3}{c}{LPIPS\,$\downarrow$} \\
\cmidrule(lr){3-5}\cmidrule(lr){6-8}
Dataset & \scriptsize{Map} & \scriptsize{P.\,Haven} & \scriptsize{LSDIR} & $\Delta$ & \scriptsize{P.\,Haven} & \scriptsize{LSDIR} & $\Delta$ \\
\midrule
\multirow{3}{*}{\textsc{Hangar}}
 & D & 36.888 & \textbf{37.529} & $+0.640$ & 0.0754 & \textbf{0.0664} & $-0.0090$ \\
 & N  & \textbf{37.687} & 37.167 & $-0.519$ & \textbf{0.0226} & 0.0353 & $+0.0127$ \\
 & A     & 36.705 & \textbf{37.389} & $+0.683$ & 0.0413 & \textbf{0.0405} & $-0.0008$ \\
\midrule
\multirow{3}{*}{\textsc{Victorian}}
 & D & 37.373 & \textbf{37.580} & $+0.206$ & 0.1024 & \textbf{0.0959} & $-0.0066$ \\
 & N  & 37.694 & \textbf{39.855} & $+2.162$ & \textbf{0.0244} & 0.0342 & $+0.0098$ \\
 & A     & 40.746 & \textbf{41.392} & $+0.646$ & 0.0455 & \textbf{0.0393} & $-0.0061$ \\
\midrule
\multirow{3}{*}{\textsc{Museum}}
 & D & 40.387 & \textbf{40.881} & $+0.494$ & 0.0521 & \textbf{0.0500} & $-0.0022$ \\
 & N  & 38.481 & \textbf{41.345} & $+2.864$ & \textbf{0.0199} & 0.0202 & $+0.0002$ \\
 & A     & 38.391 & \textbf{39.738} & $+1.346$ & 0.0311 & \textbf{0.0296} & $-0.0015$ \\
\midrule
\multirow{3}{*}{\shortstack[l]{\textsc{P.\,Haven}\\held-out}}
 & D & \textbf{32.421} & 32.408 & $-0.013$ & 0.2403 & \textbf{0.2322} & $-0.0081$ \\
 & N  & \textbf{34.766} & 33.437 & $-1.329$ & \textbf{0.0723} & 0.0955 & $+0.0232$ \\
 & A     & \textbf{35.973} & 35.213 & $-0.761$ & \textbf{0.0611} & 0.0663 & $+0.0051$ \\
\bottomrule
\end{tabular}
\end{table}

Table~\ref{tab:natural-images} shows the result split by map type, which is essential: a
single average over the three types hides the effect entirely. On \textbf{diffuse} maps
the photograph-trained model is better on LPIPS on all four test datasets, by $0.0022$ to
$0.0090$; on \textbf{ARM} maps it is better on three of four. On \textbf{normal} maps it
is better on none. The direction is consistent across scenes and an order of magnitude
above the seed spread.

\begin{table}[t]
\caption{Held-out \textsc{Poly Haven} split, per map type. On normal maps the
photograph-trained model is barely distinguishable from Lanczos upscaling, while the
texture-trained model gains $1.38$~dB over it.}
\label{tab:heldout-dna}
\small
\setlength{\tabcolsep}{4pt}
\begin{tabular}{lrrr@{\hskip 10pt}rrr}
\toprule
 & \multicolumn{3}{c}{PSNR\,$\uparrow$} & \multicolumn{3}{c}{LPIPS\,$\downarrow$} \\
\cmidrule(lr){2-4}\cmidrule(lr){5-7}
Method & diff. & norm. & ARM & diff. & norm. & ARM \\
\midrule
Bilinear            & 31.179 & 32.461 & 33.915 & 0.3962 & 0.1907 & 0.1414 \\
Bicubic             & 31.990 & 33.244 & 35.119 & 0.3694 & 0.1617 & 0.1081 \\
Lanczos             & 32.123 & 33.382 & 35.326 & 0.3833 & 0.1597 & 0.1053 \\
\midrule
Ours, LSDIR         & 32.408 & 33.437 & 35.213 & \textbf{0.2322} & 0.0955 & 0.0663 \\
Ours, \textsc{P.\,Haven} & \textbf{32.421} & \textbf{34.766} & \textbf{35.973} & 0.2403 & \textbf{0.0723} & \textbf{0.0611} \\
\bottomrule
\end{tabular}
\end{table}

Table~\ref{tab:heldout-dna} makes the failure concrete. On the held-out normal maps the
photograph-trained model reaches $33.437$~dB against $33.382$~dB for Lanczos---a gain of
$0.06$~dB---whereas the texture-trained model reaches $34.766$~dB, a gain of $1.38$~dB.
On that class, photographs teach the model almost nothing.

This is consistent with what a normal map is. Diffuse albedo and the packed greyscale
channels of an ARM map are photograph-like signals: they have edges, shading gradients and
natural colour correlations, and an operator learned from photographs transfers to them.
A normal map is not an image but a field of unit vectors encoded into RGB, concentrated
around $(0.5, 0.5, 1)$ and piecewise smooth, with none of those statistics. The
"generalized local operator" interpretation of our method therefore holds for
photograph-like channels and breaks for encoded geometric data.

We note one caveat. On \textsc{Victorian} and \textsc{Museum} normal maps, PSNR ranks the
photograph-trained model far higher ($+2.16$ and $+2.86$~dB) while LPIPS ranks it lower.
The consistent reading is that photographs bias the model towards smoother normals, which
raises PSNR against a smooth reference and costs perceptual detail; this is the same
effect discussed in Sec.~\ref{sec:comparison} and a further reason to read LPIPS.

\paragraph{Dataset size.}
Holding the optimization budget fixed and varying only the number of unique training
texels, test PSNR on \textsc{Hangar} rises from $37.463$~dB at $0.22$~Gpixel (182 images)
to $37.678$~dB at $3.56$~Gpixel (3236 images): a $16\times$ increase in unique data is
worth $+0.215$~dB. The $1.78$~Gpixel \textsc{Poly Haven} dataset already sits on the flat
part of this curve. The limiting factor is not how much training data is available but
which map types it covers.

\subsubsection{Model Capacity}
\label{sec:capacity}

If the network is evaluated inside the texture sampler, its parameter count and its
texel fetch count decide whether it is deployable. Here, we test how far it can be
reduced. We compare the default configuration ($d=256$, $L=4$ residual
blocks, $536{,}588$ parameters at a $3\times3$ window) against two networks with no
residual blocks and a single hidden-to-hidden layer, i.e. \mbox{$27 \to d \to d \to 12$}
with $d=64$ ($6{,}732$ parameters) and $d=32$ ($2{,}348$ parameters). The analytic
upscaling path and the residual formulation are unchanged.

\begin{table}[t]
\caption{Model capacity on \textsc{Hangar}, per map type, at a $3\times3$ window
($2.25$ texel fetches per output pixel for every row). Even a $2{,}348$-parameter
network---$228\times$ smaller than the one used elsewhere in this paper---improves on
Lanczos upscaling on both metrics.}
\label{tab:capacity}
\small
\setlength{\tabcolsep}{4pt}
\begin{tabular}{lrrrr@{\hskip 8pt}rrr}
\toprule
 & & \multicolumn{3}{c}{PSNR\,$\uparrow$} & \multicolumn{3}{c}{LPIPS\,$\downarrow$} \\
\cmidrule(lr){3-5}\cmidrule(lr){6-8}
Model & Params & diff. & norm. & ARM & diff. & norm. & ARM \\
\midrule
\scriptsize{Bilinear} & --- & 34.305 & 35.311 & 35.297 & 0.1930 & 0.0879 & 0.0943 \\
\scriptsize{Bicubic}  & --- & 35.425 & 36.326 & 36.760 & 0.1562 & 0.0669 & 0.0679 \\
\scriptsize{Lanczos}  & --- & 35.639 & 36.521 & 37.045 & 0.1573 & 0.0647 & 0.0652 \\
\midrule
\scriptsize{$32\to32$}   & 2{,}348   & 36.675 & 36.457 & 36.819 & 0.0766 & 0.0386 & 0.0446 \\
\scriptsize{$64\to64$}   & 6{,}732   & 36.987 & 36.766 & 37.207 & 0.0731 & 0.0412 & 0.0423 \\
\scriptsize{$d{=}256$, $L{=}4$} & 536{,}588 & \textbf{37.553} & \textbf{37.304} & \textbf{37.522} & \textbf{0.0665} & \textbf{0.0324} & \textbf{0.0406} \\
\bottomrule
\end{tabular}
\end{table}

Table~\ref{tab:capacity} shows that every network tested improves on all three analytic
filters, on both metrics and on all three map types. The smallest, at $2{,}348$
parameters, is $228\times$ smaller than the configuration used elsewhere in this paper and
still improves on Lanczos by $1.04$~dB on diffuse maps while reducing diffuse LPIPS from
$0.1573$ to $0.0766$, a $51\%$ reduction, at the same $2.25$ texel fetches per output
pixel. Growing the network from $2{,}348$ to $536{,}588$ parameters---a $228\times$
increase---adds a further $0.88$~dB on diffuse. Most of the quality available to this
method is therefore reached at the cheap end of the capacity range, which is the
range a shader can afford.

The ordering is stable: across \textsc{Hangar}, \textsc{Victorian}, \textsc{Museum} and
the \textsc{Poly Haven} test split, the three capacities rank in the same order on seven
of the eight scene--metric combinations, the exception being LPIPS on \textsc{Museum},
where the two smaller networks are tied to within $10^{-5}$.

\paragraph{Capacity interacts with the input window.}
Enlarging the input patch from $3\times3$ to $5\times5$ ($m=2$, $75$ inputs, $6.25$ texel
fetches per output pixel instead of $2.25$) is only worthwhile for the large network.

\begin{table}[t]
\caption{Effect of enlarging the input window from $3\times3$ to $5\times5$ on
\textsc{Hangar}, by model size and map type. Entries are $5\times5$ minus $3\times3$, so a
positive $\Delta$PSNR and a negative $\Delta$LPIPS are improvements. Only the largest
network improves on both metrics, and the gain is concentrated on diffuse maps.}
\label{tab:window-capacity}
\small
\setlength{\tabcolsep}{4pt}
\begin{tabular}{llrrrr}
\toprule
Model & Metric & \scriptsize{diff.} & \scriptsize{norm.} & \scriptsize{ARM} & \scriptsize{all} \\
\midrule
\multirow{2}{*}{\scriptsize{$32\to32$}}
 & $\Delta$PSNR\,$\uparrow$   & $+0.323$ & $-0.006$ & $+0.018$ & $+0.112$ \\
 & $\Delta$LPIPS\,$\downarrow$ & $+0.0039$ & $+0.0037$ & $+0.0013$ & $+0.0030$ \\
\midrule
\multirow{2}{*}{\scriptsize{$64\to64$}}
 & $\Delta$PSNR\,$\uparrow$   & $+0.425$ & $-0.021$ & $-0.174$ & $+0.076$ \\
 & $\Delta$LPIPS\,$\downarrow$ & $+0.0006$ & $\mathbf{-0.0030}$ & $+0.0020$ & $-0.0001$ \\
\midrule
\multirow{2}{*}{\scriptsize{$d{=}256$, $L{=}4$}}
 & $\Delta$PSNR\,$\uparrow$   & $\mathbf{+0.660}$ & $\mathbf{+0.131}$ & $-0.099$ & $\mathbf{+0.229}$ \\
 & $\Delta$LPIPS\,$\downarrow$ & $\mathbf{-0.0042}$ & $-0.0028$ & $+0.0011$ & $\mathbf{-0.0019}$ \\
\bottomrule
\end{tabular}
\end{table}

Table~\ref{tab:window-capacity} shows why. The $d{=}256$ model gains $+0.229$~dB and
$-0.0019$ LPIPS overall, whereas the $32\to32$ model gains $+0.112$~dB but \emph{loses}
$0.0030$ LPIPS, and the $64\to64$ model is unchanged on LPIPS ($-0.0001$). A small network
cannot exploit the additional 48 inputs and spends capacity representing them rather than
using them. The gain is also concentrated on one map type: for the $d{=}256$ model the
change is worth $+0.660$~dB on diffuse but only $+0.131$~dB on normal and $-0.099$~dB on
ARM, and no model improves on ARM at all. Since the wider window costs $2.8\times$ as many
texel fetches, $3\times3$ is both the better and the cheaper choice for a shader-sized
network.

\subsubsection{Implications}
\label{sec:data-capacity-discussion}

Two practical conclusions follow. First, the training dataset does not need to consist of
PBR materials for photograph-like channels, which removes the main data-collection
obstacle to training a generalized upscaler; but it must cover encoded channels such as
normal maps explicitly, because photographs do not teach them. A dataset combining natural
images for diffuse and ARM maps with PBR normal maps is the obvious construction and we
leave its evaluation to future work. Second, the network can be made dramatically smaller
than the configuration evaluated in Sec.~\ref{sec:comparison} while still improving on the
analytic filters it replaces, which strengthens the case for executing the model at
texture-sampling time rather than as a preprocessing step.
\section{Conclusion}

We have presented a lightweight neural texture upscaling method that operates at texture sampling time without requiring any modifications to existing texture assets or content creation pipelines. A single generalized MLP, trained once on a diverse collection of PBR texture patches, can be substituted directly for the conventional analytic texture sampler — enabling higher-quality rendering in existing scenes and games simply by swapping the sampling step. This stands in contrast to prior neural approaches such as NTC and NTBC, which require per-asset optimization and introduce content creation overhead, as well as offline super-resolution methods such as PFT, which cannot be executed at sampling time due to their computational cost and lack of random access.         

Our evaluation across multiple texture classes and rendering scenes demonstrates consistent improvements in perceptual quality over analytic upscaling methods. The proposed method outperforms bilinear, bicubic, and Lanczos interpolation in LPIPS across all tested datasets, and approaches PFT quality in several cases despite being orders of magnitude smaller and designed for real-time execution. We also showed that a single unified model trained on all three texture classes (diffuse, normal, ARM) achieves favorable balanced results, further simplifying deployment. When per-asset quality is prioritized, the model can be overfit to a specific texture set with consistent gains at the cost of additional preprocessing.

Future work could explore incorporating texture-class-aware priors, extending to higher upscale ratios, or combining our generalized model with lightweight per-asset adaptation to close the quality gap with overfitted methods while preserving the pipeline advantages of the generalized approach. An important direction not addressed in this work is a systematic runtime performance evaluation of the MLP shader integrated into a practical rendering pipeline. Characterizing the per-sample inference cost, its impact on overall frame time, and the trade-off between model size and rendering throughput would be essential for practical deployment in games and real-time applications. 

\section*{Acknowledgement}
\textsc{Hangar} is our original dataset. \textsc{Victorian} and \textsc{Museum} datasets are from \url{https://www.fab.com/listings/f6d06d4c-251a-464b-aaf0-3e68c8503582}, and \url{https://www.fab.com/listings/32aca4d1-84c0-4ad9-950c-9dde7995dca8}. Authors like to thank Anton Timoshenko and Aleksandr Nefedov for their help with the development.

%
%
%
%

\bibliographystyle{ACM-Reference-Format}
\bibliography{main}

\clearpage
\onecolumn
\appendix
\section{Supplemental Material}
\begin{table*}[ht]
  \centering
  \caption{LPIPS per texture in the diffuse class in the \textsc{Poly Haven} dataset}
  \small
  \begin{tabular}{lrrrrr}
    \toprule
    Texture & baseline & bilinear & bicubic & lanczos3 & PFT \\
    \midrule
    castle\_brick\_02\_white\_diff\_1k & 0.1607 & 0.4271 & 0.3949 & 0.4273 & 0.2235 \\
    blue\_floor\_tiles\_01\_diff\_1k & 0.1209 & 0.2719 & 0.2503 & 0.2640 & 0.1449 \\
    brushed\_concrete\_2\_diff\_1k & 0.2318 & 0.3555 & 0.3351 & 0.3444 & 0.2743 \\
    gravel\_concrete\_diff\_1k & 0.5256 & 0.7272 & 0.6714 & 0.6929 & 0.3920 \\
    floral\_jacquard\_diff\_1k & 0.4385 & 0.5948 & 0.5659 & 0.5701 & 0.4196 \\
    denim\_fabric\_diff\_1k & 0.5518 & 0.8128 & 0.7742 & 0.7793 & 0.4409 \\
    floor\_tiles\_08\_diff\_1k & 0.1639 & 0.2931 & 0.2630 & 0.2700 & 0.1564 \\
    cobblestone\_large\_01\_diff\_1k & 0.1785 & 0.3515 & 0.3243 & 0.3429 & 0.1939 \\
    green\_metal\_rust\_diff\_1k & 0.0839 & 0.2158 & 0.1525 & 0.1434 & 0.0520 \\
    metal\_plate\_02\_diff\_1k & 0.1893 & 0.3209 & 0.3030 & 0.3154 & 0.2089 \\
    volcanic\_herringbone\_01\_diff\_1k & 0.2172 & 0.3528 & 0.3163 & 0.3214 & 0.2209 \\
    rock\_wall\_06\_diff\_1k & 0.1569 & 0.3206 & 0.2994 & 0.3160 & 0.1586 \\
    roof\_slates\_02\_diff\_1k & 0.1185 & 0.2360 & 0.2660 & 0.3003 & 0.0806 \\
    roof\_09\_diff\_1k & 0.0968 & 0.2324 & 0.2065 & 0.2196 & 0.1089 \\
    dark\_planks\_diff\_1k & 0.1637 & 0.2828 & 0.2524 & 0.2549 & 0.1504 \\
    brown\_planks\_03\_diff\_1k & 0.3378 & 0.4385 & 0.4399 & 0.4645 & 0.3504 \\
    rocks\_ground\_06\_diff\_1k & 0.3244 & 0.5019 & 0.4652 & 0.4899 & 0.3008 \\
    \bottomrule
  \end{tabular}
\end{table*}
\begin{table*}[ht]
  \centering
  \caption{LPIPS per texture in the diffuse class in the \textsc{Hanger} dataset}
  \small
  \begin{tabular}{lrrrrr}
    \toprule
    Texture & baseline & bilinear & bicubic & lanczos3 & PFT \\
    \midrule
    Arch\_Panels\_dark\_D & 0.0699 & 0.2247 & 0.1928 & 0.1977 & 0.0259 \\
    Barrel\_2\_Base\_color & 0.0247 & 0.1035 & 0.0842 & 0.0830 & 0.0109 \\
    Base\_001\_basecolor & 0.0901 & 0.2328 & 0.1996 & 0.2085 & 0.0515 \\
    Base\_002\_basecolor & 0.1174 & 0.2746 & 0.2417 & 0.2519 & 0.0741 \\
    Base\_003\_basecolor & 0.0445 & 0.1142 & 0.0900 & 0.0899 & 0.0306 \\
    BigBox\_color & 0.0959 & 0.2097 & 0.1878 & 0.1938 & 0.0411 \\
    Box\_03\_remake1k\_1001\_BaseColor\_001 & 0.0327 & 0.1111 & 0.0798 & 0.0769 & 0.0103 \\
    Box\_03\_remake1k\_1001\_BaseColor\_002 & 0.0240 & 0.0826 & 0.0564 & 0.0532 & 0.0084 \\
    Box\_03\_remake1k\_1001\_BaseColor\_003 & 0.0352 & 0.1233 & 0.0944 & 0.0934 & 0.0121 \\
    Ceiling\_dark\_diffuse & 0.1436 & 0.2731 & 0.2464 & 0.2545 & 0.0886 \\
    Cloth\_Base\_color & 0.1856 & 0.3531 & 0.3268 & 0.3462 & 0.1914 \\
    Cruser\_colour2\_light & 0.0755 & 0.2306 & 0.2102 & 0.2214 & 0.0275 \\
    DroneWarriorTexRebild\_Color & 0.0873 & 0.2426 & 0.2017 & 0.2030 & 0.0288 \\
    Entrance\_gates\_Base\_color & 0.0893 & 0.2350 & 0.2073 & 0.2155 & 0.0359 \\
    Floor\_001\_basecolor & 0.0486 & 0.1545 & 0.0927 & 0.0853 & 0.0370 \\
    Floor\_hatch\_Base\_color & 0.0842 & 0.2048 & 0.1701 & 0.1712 & 0.0466 \\
    ForTex2\_lambert1\_BaseColor & 0.0839 & 0.2206 & 0.1609 & 0.1556 & 0.0271 \\
    Gates\_Base\_color & 0.0781 & 0.2016 & 0.1606 & 0.1600 & 0.0426 \\
    Hangar-Decals\_BaseColor & 0.0359 & 0.1238 & 0.0902 & 0.0876 & 0.0168 \\
    Hangar-FloorTile-v1\_BaseColor & 0.1116 & 0.2539 & 0.2061 & 0.2051 & 0.0919 \\
    Hangar-Wall-LP\_BaseColor\_light & 0.1030 & 0.2033 & 0.1756 & 0.1797 & 0.0531 \\
    HangarDoorFrame-LP\_Hangar-Tunnel\_\_BaseColor & 0.0693 & 0.1940 & 0.1535 & 0.1533 & 0.0540 \\
    HatchBake\_BaseColor & 0.0389 & 0.1203 & 0.0900 & 0.0868 & 0.0288 \\
    Low\_initialShadingGroup1\_2k\_BaseColor & 0.0760 & 0.2056 & 0.1615 & 0.1606 & 0.0467 \\
    LuxSimpleMetal\_basecolor & 0.0694 & 0.2382 & 0.1473 & 0.1282 & 0.0590 \\
    MI\_Trims\_B\_Black\_BaseColor & 0.0847 & 0.1987 & 0.1566 & 0.1516 & 0.0516 \\
    New\_Graph\_BaseColor & 0.0427 & 0.1359 & 0.0988 & 0.0990 & 0.0332 \\
    ResinFloorSimple\_basecolor & 0.0377 & 0.2647 & 0.0998 & 0.0817 & 0.0208 \\
    ResinFloorSimple\_Stripes\_basecolor & 0.0104 & 0.0503 & 0.0266 & 0.0228 & 0.0057 \\
    SciFiBrick\_BaseColor & 0.0341 & 0.1014 & 0.0767 & 0.0750 & 0.0124 \\
    Small\_ship\_Base\_color & 0.1439 & 0.3084 & 0.2844 & 0.3025 & 0.0605 \\
    T\_Terminal\_UI & 0.0779 & 0.1555 & 0.1334 & 0.1331 & 0.0264 \\
    Tank2\_LP\_1001\_BaseColor & 0.0611 & 0.1292 & 0.1035 & 0.1028 & 0.0380 \\
    Trim\_2\_lp\_Trim2\_BaseColor & 0.0711 & 0.2111 & 0.1663 & 0.1647 & 0.0366 \\
    Trim\_Tile\_lambert1\_BaseColor & 0.0203 & 0.0691 & 0.0475 & 0.0459 & 0.0174 \\
    Trim\_block\_lp\_TrimBlock\_BaseColor & 0.0614 & 0.1768 & 0.1418 & 0.1431 & 0.0262 \\
    Wall\_dark\_diffuse & 0.1301 & 0.2838 & 0.2492 & 0.2556 & 0.0660 \\
    Weapon\_D & 0.0974 & 0.2594 & 0.2390 & 0.2523 & 0.0304 \\
    barrier\_lp\_Barrier\_BaseColor & 0.0475 & 0.1311 & 0.0918 & 0.0872 & 0.0258 \\
    spaceship\_body\_3\_Color & 0.0655 & 0.1872 & 0.1545 & 0.1578 & 0.0322 \\
    spaceship\_engines\_3\_Color & 0.0591 & 0.1805 & 0.1453 & 0.1478 & 0.0309 \\
    spaceship\_wings\_2\_Color & 0.0783 & 0.2070 & 0.1724 & 0.1752 & 0.0373 \\
    tank3\_base\_color & 0.1754 & 0.3004 & 0.2691 & 0.2718 & 0.1390 \\
    tank\_4\_base\_color & 0.1267 & 0.2561 & 0.2151 & 0.2153 & 0.0765 \\
    wire\_albedo & 0.0473 & 0.1469 & 0.1282 & 0.1327 & 0.0068 \\
    \bottomrule
  \end{tabular}
\end{table*}

\begin{table*}[ht]
  \centering
  \caption{LPIPS per texture in the normal class in the \textsc{Hanger} dataset}
  \small
  \begin{tabular}{lrrrrr}
    \toprule
    Texture & baseline & bilinear & bicubic & lanczos3 & PFT \\
    \midrule
    Arch\_Panels\_N & 0.0143 & 0.0937 & 0.0702 & 0.0678 & 0.0055 \\
    Barrel\_2\_Normal & 0.0130 & 0.0626 & 0.0430 & 0.0411 & 0.0039 \\
    Base\_001\_normal & 0.0196 & 0.0803 & 0.0634 & 0.0647 & 0.0093 \\
    Base\_002\_normal & 0.0556 & 0.1637 & 0.1250 & 0.1192 & 0.0471 \\
    Base\_003\_normal & 0.0335 & 0.0903 & 0.0771 & 0.0768 & 0.0275 \\
    BigBox\_normal & 0.0222 & 0.0888 & 0.0680 & 0.0656 & 0.0163 \\
    Box\_03\_remake1k\_1001\_Normal & 0.0178 & 0.0508 & 0.0358 & 0.0324 & 0.0130 \\
    Ceiling\_normal & 0.0336 & 0.1161 & 0.0890 & 0.0862 & 0.0204 \\
    Cloth\_Normal & 0.0042 & 0.0392 & 0.0226 & 0.0204 & 0.0019 \\
    Cruser\_normals & 0.0333 & 0.1459 & 0.1174 & 0.1149 & 0.0139 \\
    DroneWarriorTexRebild\_Normal & 0.0291 & 0.1202 & 0.0889 & 0.0849 & 0.0136 \\
    Entrance\_gates\_Normal & 0.0111 & 0.0688 & 0.0459 & 0.0428 & 0.0074 \\
    Floor\_001\_normal & 0.0230 & 0.0726 & 0.0590 & 0.0582 & 0.0198 \\
    Floor\_hatch\_Normal & 0.0213 & 0.1137 & 0.0871 & 0.0836 & 0.0081 \\
    ForTex2\_lambert1\_Normal & 0.0071 & 0.0517 & 0.0348 & 0.0331 & 0.0032 \\
    Gates\_Normal & 0.0256 & 0.1349 & 0.1044 & 0.1005 & 0.0086 \\
    Hangar-Decals\_Normal & 0.0424 & 0.1518 & 0.1048 & 0.0962 & 0.0373 \\
    Hangar-FloorTile-v1\_Normal & 0.0869 & 0.2731 & 0.2249 & 0.2223 & 0.0417 \\
    Hangar-Wall-LP\_Normal\_for\_logo & 0.0446 & 0.1606 & 0.1271 & 0.1250 & 0.0287 \\
    HangarDoorFrame-LP\_Hangar-Tunnel\_\_Normal & 0.0216 & 0.1067 & 0.0713 & 0.0670 & 0.0091 \\
    HatchBake\_Normal & 0.0141 & 0.0677 & 0.0525 & 0.0512 & 0.0068 \\
    Low\_initialShadingGroup1\_2k\_Normal & 0.0623 & 0.1769 & 0.1404 & 0.1357 & 0.0376 \\
    LuxSimpleMetal\_normal & 0.0005 & 0.0008 & 0.0006 & 0.0006 & 0.0009 \\
    MI\_Trims\_B\_Normal & 0.0222 & 0.0800 & 0.0614 & 0.0592 & 0.0200 \\
    New\_Graph\_Normal & 0.0069 & 0.0302 & 0.0225 & 0.0211 & 0.0047 \\
    ResinFloorSimple\_normal & 0.0050 & 0.0266 & 0.0181 & 0.0167 & 0.0040 \\
    ResinFloorSimple\_Stripes\_normal & 0.0067 & 0.0313 & 0.0218 & 0.0203 & 0.0048 \\
    SciFiBrick\_Normal & 0.0081 & 0.0510 & 0.0349 & 0.0332 & 0.0039 \\
    Small\_ship\_Normal & 0.0275 & 0.1535 & 0.1288 & 0.1297 & 0.0071 \\
    T\_Metal\_Scratches\_M\_normal & 0.0006 & 0.0011 & 0.0008 & 0.0008 & 0.0018 \\
    Table\_Normal & 0.0028 & 0.0167 & 0.0101 & 0.0097 & 0.0020 \\
    Tank2\_LP\_1001\_Normal & 0.0185 & 0.0384 & 0.0287 & 0.0272 & 0.0148 \\
    Trim\_2\_lp\_Trim2\_Normal & 0.0377 & 0.1111 & 0.0922 & 0.0917 & 0.0222 \\
    Trim\_Tile\_lambert1\_Normal & 0.0070 & 0.0193 & 0.0155 & 0.0152 & 0.0075 \\
    Trim\_block\_lp\_TrimBlock\_Normal & 0.0348 & 0.1114 & 0.0887 & 0.0864 & 0.0260 \\
    Wall\_normal & 0.0452 & 0.1641 & 0.1248 & 0.1212 & 0.0218 \\
    Weapon\_N & 0.0179 & 0.1048 & 0.0751 & 0.0713 & 0.0037 \\
    barrier\_lp\_Barrier\_Normal & 0.0337 & 0.1067 & 0.0795 & 0.0761 & 0.0224 \\
    spaceship\_body\_3\_Normal & 0.0179 & 0.0784 & 0.0588 & 0.0570 & 0.0093 \\
    spaceship\_engines\_3\_Normal & 0.0082 & 0.0353 & 0.0259 & 0.0249 & 0.0060 \\
    spaceship\_wings\_2\_Normal & 0.0235 & 0.1020 & 0.0776 & 0.0751 & 0.0131 \\
    tank3\_normal & 0.0078 & 0.0410 & 0.0278 & 0.0263 & 0.0043 \\
    tank\_4\_normal & 0.0116 & 0.0568 & 0.0418 & 0.0402 & 0.0074 \\
    wire\_normal & 0.0170 & 0.0751 & 0.0556 & 0.0548 & 0.0061 \\
    \bottomrule
  \end{tabular}
\end{table*}

\begin{table*}[ht]
  \centering
  \caption{LPIPS per texture in the ARM class in the \textsc{Hanger} dataset}
  \small
  \begin{tabular}{lrrrrr}
    \toprule
    Texture & baseline & bilinear & bicubic & lanczos3 & PFT \\
    \midrule
    Arch\_Panels\_RoughMet\_D & 0.0502 & 0.0987 & 0.0752 & 0.0721 & 0.0252 \\
    Barrel\_2\_ORM & 0.0124 & 0.0425 & 0.0278 & 0.0261 & 0.0047 \\
    Base\_001\_ORM & 0.0250 & 0.0938 & 0.0580 & 0.0533 & 0.0135 \\
    Base\_002\_ORM & 0.0400 & 0.1338 & 0.0880 & 0.0845 & 0.0211 \\
    Base\_003\_ORM & 0.0130 & 0.0397 & 0.0322 & 0.0321 & 0.0124 \\
    BigBox\_ORM & 0.0193 & 0.0484 & 0.0294 & 0.0273 & 0.0088 \\
    Box\_03\_remake1k\_1001\_ORM & 0.0079 & 0.0290 & 0.0151 & 0.0138 & 0.0038 \\
    Ceiling\_RoughMet & 0.0910 & 0.1719 & 0.1419 & 0.1420 & 0.0532 \\
    Cloth\_Roughness\_green\_channel & 0.0113 & 0.0338 & 0.0213 & 0.0204 & 0.0065 \\
    DroneWarriorTexRebild\_RoughMet & 0.0314 & 0.0775 & 0.0567 & 0.0537 & 0.0200 \\
    Entrance\_gates\_Roughness\_Metalness & 0.0367 & 0.0969 & 0.0654 & 0.0617 & 0.0240 \\
    Floor\_001\_ORM & 0.0301 & 0.0925 & 0.0525 & 0.0473 & 0.0128 \\
    Floor\_hatch\_Roughness\_Metalness & 0.0318 & 0.0839 & 0.0596 & 0.0562 & 0.0222 \\
    ForTex2\_lambert1\_RoughMet & 0.0454 & 0.1124 & 0.0684 & 0.0621 & 0.0193 \\
    Gates\_Roughness\_Metalness & 0.0108 & 0.0436 & 0.0272 & 0.0254 & 0.0049 \\
    Hangar-Decals\_Roughness\_Metalness & 0.1080 & 0.2225 & 0.1747 & 0.1674 & 0.0628 \\
    Hangar-FloorTile-v1\_ORM & 0.0447 & 0.1103 & 0.0797 & 0.0760 & 0.0173 \\
    Hangar-Wall-LP\_Roughness\_Metalnessfor\_logo & 0.0415 & 0.1064 & 0.0734 & 0.0682 & 0.0277 \\
    HangarDoorFrame-LP\_Hangar-Tunnel\_\_Roughness\_Metalness & 0.0133 & 0.0447 & 0.0247 & 0.0222 & 0.0044 \\
    HatchBake\_ORM & 0.0848 & 0.1633 & 0.1266 & 0.1201 & 0.0490 \\
    Low\_initialShadingGroup1\_2k\_ORM & 0.0458 & 0.1024 & 0.0736 & 0.0699 & 0.0234 \\
    LuxSimpleMetal\_roughness2\_green\_channel & 0.0498 & 0.1172 & 0.0824 & 0.0789 & 0.0251 \\
    MI\_Trims\_B\_Roughness\_green\_channel & 0.0045 & 0.0101 & 0.0062 & 0.0058 & 0.0027 \\
    New\_Graph\_ORM & 0.0080 & 0.0309 & 0.0211 & 0.0201 & 0.0042 \\
    ResinFloorSimple\_ORM & 0.0226 & 0.1375 & 0.0545 & 0.0466 & 0.0041 \\
    ResinFloorSimple\_Stripes\_ORM & 0.0186 & 0.0938 & 0.0402 & 0.0348 & 0.0052 \\
    SciFiBrick\_ORM & 0.0170 & 0.0544 & 0.0348 & 0.0329 & 0.0070 \\
    ShinyPlastic\_Roughness\_green\_channel & 0.0093 & 0.0311 & 0.0193 & 0.0184 & 0.0052 \\
    Small\_ship\_Roughness\_Metalness & 0.1604 & 0.3013 & 0.2817 & 0.2914 & 0.1370 \\
    T\_Metal\_Scratches\_M8\_green\_chanel-1 & 0.1446 & 0.2329 & 0.1781 & 0.1715 & 0.0734 \\
    T\_Metal\_Scratches\_M8\_green\_chanel & 0.0801 & 0.1226 & 0.1022 & 0.0999 & 0.0643 \\
    T\_Metal\_Scratches\_M\_green\_channel & 0.1832 & 0.3072 & 0.2452 & 0.2389 & 0.1171 \\
    Table\_Roughness\_green\_channel & 0.0049 & 0.0209 & 0.0121 & 0.0110 & 0.0027 \\
    Tank2\_LP\_1001\_Roughness\_Metalness & 0.0137 & 0.0320 & 0.0225 & 0.0213 & 0.0089 \\
    Trim\_2\_lp\_Trim2\_OcclusionRoughnessMetallic & 0.0249 & 0.0662 & 0.0489 & 0.0466 & 0.0132 \\
    Trim\_Tile\_lambert1\_OcclusionRoughnessMetallic & 0.0030 & 0.0101 & 0.0059 & 0.0056 & 0.0021 \\
    Trim\_block\_lp\_TrimBlock\_OcclusionRoughnessMetallic & 0.0194 & 0.0528 & 0.0377 & 0.0353 & 0.0100 \\
    Wall\_RoughMet & 0.0912 & 0.1754 & 0.1455 & 0.1438 & 0.0625 \\
    Weapon\_RoughMet & 0.0208 & 0.0627 & 0.0408 & 0.0379 & 0.0121 \\
    barrier\_lp\_Barrier\_OcclusionRoughnessMetallic & 0.0285 & 0.0767 & 0.0559 & 0.0534 & 0.0150 \\
    spaceship\_body\_3\_RM & 0.0132 & 0.0545 & 0.0339 & 0.0321 & 0.0065 \\
    spaceship\_engines\_3\_RM & 0.0098 & 0.0421 & 0.0248 & 0.0233 & 0.0041 \\
    spaceship\_wings\_2\_RM & 0.0208 & 0.0728 & 0.0490 & 0.0471 & 0.0102 \\
    tank3\_Orm & 0.0346 & 0.0750 & 0.0524 & 0.0495 & 0.0230 \\
    tank\_4\_ORM & 0.0625 & 0.1157 & 0.0924 & 0.0901 & 0.0332 \\
    wireRough\_Roughness\_Metalness & 0.0283 & 0.0939 & 0.0634 & 0.0596 & 0.0142 \\
    \bottomrule
  \end{tabular}
\end{table*}
\begin{table*}
  \centering
  \caption{LPIPS per texture in the diffuse class in the \textsc{Victorian} dataset}
  \small
  \begin{tabular}{lrrrrr}
    \toprule
    Texture & Ours & Bilinear & Bicubic & Lanczos3 & PFT \\
    \midrule
    T\_ArchHalf\_A         & 0.1340 & 0.2823 & 0.2487 & 0.2549 & 0.1127 \\
    T\_ArchLarge\_A        & 0.0937 & 0.2400 & 0.1861 & 0.1857 & 0.0917 \\
    T\_BaseFloor\_A        & 0.1034 & 0.2195 & 0.1922 & 0.1964 & 0.1404 \\
    T\_Bench\_A            & 0.0642 & 0.1359 & 0.1216 & 0.1271 & 0.0324 \\
    T\_Bin\_A              & 0.1236 & 0.2266 & 0.1967 & 0.1982 & 0.0935 \\
    T\_Brick\_A            & 0.1031 & 0.2410 & 0.2112 & 0.2201 & 0.1208 \\
    T\_Ceiling\_A          & 0.0813 & 0.1874 & 0.1590 & 0.1623 & 0.0845 \\
    T\_Clock\_A            & 0.0489 & 0.1536 & 0.1167 & 0.1170 & 0.0238 \\
    T\_ControlGate\_A      & 0.0386 & 0.1246 & 0.0823 & 0.0768 & 0.0202 \\
    T\_Door\_A             & 0.0892 & 0.2084 & 0.1703 & 0.1677 & 0.0861 \\
    T\_FloorBorderSmall\_A & 0.0896 & 0.2019 & 0.1645 & 0.1624 & 0.0898 \\
    T\_FloorBorder\_A      & 0.0419 & 0.1284 & 0.0924 & 0.0947 & 0.0429 \\
    T\_FloorChecker\_A     & 0.0613 & 0.1858 & 0.1360 & 0.1339 & 0.0444 \\
    T\_FloorTilesDetails\_A & 0.0963 & 0.2217 & 0.2084 & 0.2167 & 0.0289 \\
    T\_Ground\_A           & 0.1547 & 0.3504 & 0.2809 & 0.2802 & 0.1886 \\
    T\_InfoTable\_A        & 0.0793 & 0.1824 & 0.1538 & 0.1584 & 0.0675 \\
    T\_Light\_A            & 0.1766 & 0.2862 & 0.2523 & 0.2541 & 0.1580 \\
    T\_Paintings\_A        & 0.0710 & 0.2430 & 0.1831 & 0.1842 & 0.0744 \\
    T\_PhoneBooth\_A       & 0.0619 & 0.1631 & 0.1347 & 0.1361 & 0.0385 \\
    T\_PillarFrame\_A      & 0.0986 & 0.2137 & 0.1889 & 0.1914 & 0.0798 \\
    T\_PillarLarge\_A      & 0.0948 & 0.2203 & 0.1803 & 0.1796 & 0.0981 \\
    T\_Pillar\_A           & 0.1470 & 0.2863 & 0.2560 & 0.2601 & 0.1213 \\
    T\_Pipe\_A             & 0.4576 & 0.6305 & 0.5908 & 0.6024 & 0.4282 \\
    T\_Roof\_A             & 0.0943 & 0.2331 & 0.1862 & 0.1890 & 0.1039 \\
    T\_SecondaryFloor\_A   & 0.1358 & 0.2600 & 0.2242 & 0.2264 & 0.1074 \\
    T\_SignalLight\_A      & 0.1133 & 0.2478 & 0.2129 & 0.2212 & 0.1151 \\
    T\_Signs\_A            & 0.0681 & 0.1751 & 0.1493 & 0.1541 & 0.0274 \\
    T\_Stairs\_A           & 0.1268 & 0.2717 & 0.2371 & 0.2425 & 0.1094 \\
    T\_TicketMachine\_A    & 0.0735 & 0.1873 & 0.1570 & 0.1593 & 0.0560 \\
    T\_Tracks\_A           & 0.0819 & 0.1917 & 0.1672 & 0.1754 & 0.0796 \\
    T\_TrainBase\_A        & 0.0797 & 0.1782 & 0.1548 & 0.1582 & 0.0435 \\
    T\_TrainParts\_A       & 0.1105 & 0.2610 & 0.2315 & 0.2412 & 0.0907 \\
    T\_Wall01\_A           & 0.1171 & 0.2480 & 0.2115 & 0.2113 & 0.1360 \\
    T\_Wall02\_A           & 0.0976 & 0.2245 & 0.1830 & 0.1780 & 0.1002 \\
    T\_Wall03\_A           & 0.0615 & 0.1811 & 0.1319 & 0.1290 & 0.0609 \\
    T\_Window\_A           & 0.0492 & 0.1476 & 0.1076 & 0.1024 & 0.0328 \\
    T\_WindowsOrnament\_A  & 0.0754 & 0.1991 & 0.1518 & 0.1488 & 0.0490 \\
    \bottomrule
  \end{tabular}
\end{table*}

\begin{table*}[ht]
  \centering
  \caption{LPIPS per texture in the normal class in the \textsc{Victorian} dataset}
  \small
  \begin{tabular}{lrrrrr}
    \toprule
    Texture & Ours & Bilinear & Bicubic & Lanczos3 & PFT \\
    \midrule
    MI\_ArchLarge\_bake\_Normal        & 0.0097 & 0.0546 & 0.0364 & 0.0352 & 0.0053 \\
    MI\_Bench\_bake\_Normal            & 0.0299 & 0.1129 & 0.1047 & 0.1079 & 0.0207 \\
    MI\_Bin\_bake\_Normal              & 0.0139 & 0.0288 & 0.0186 & 0.0177 & 0.0043 \\
    MI\_Brick\_bake\_Normal            & 0.0387 & 0.1478 & 0.1371 & 0.1423 & 0.0612 \\
    MI\_Ceiling\_bake\_Normal          & 0.0177 & 0.0595 & 0.0484 & 0.0483 & 0.0164 \\
    MI\_Clock\_bake\_Normal            & 0.0112 & 0.0682 & 0.0463 & 0.0441 & 0.0068 \\
    MI\_ControlGate\_bake\_Normal      & 0.0151 & 0.0581 & 0.0431 & 0.0422 & 0.0109 \\
    MI\_Door\_bake\_Normal             & 0.0191 & 0.0885 & 0.0696 & 0.0691 & 0.0140 \\
    MI\_FloorBorder\_bake\_Normal      & 0.0229 & 0.0581 & 0.0276 & 0.0249 & 0.0075 \\
    MI\_FloorChecker\_bake\_Normal     & 0.0096 & 0.0502 & 0.0322 & 0.0312 & 0.0117 \\
    MI\_FloorTileDetails\_bake\_Normal & 0.0609 & 0.2355 & 0.1703 & 0.1634 & 0.0391 \\
    MI\_FloorTiles\_bake\_Normal       & 0.0166 & 0.0677 & 0.0585 & 0.0591 & 0.0215 \\
    MI\_HalfArch\_bake\_Normal         & 0.0074 & 0.0489 & 0.0349 & 0.0339 & 0.0057 \\
    MI\_InfoTable\_bake\_Normal        & 0.0184 & 0.0789 & 0.0561 & 0.0541 & 0.0106 \\
    MI\_Light\_bake\_Normal            & 0.0099 & 0.0325 & 0.0200 & 0.0185 & 0.0038 \\
    MI\_PhoneBooth\_bake\_Normal       & 0.0025 & 0.0165 & 0.0097 & 0.0088 & 0.0036 \\
    MI\_PillarFrame\_bake\_Normal      & 0.0135 & 0.0624 & 0.0488 & 0.0477 & 0.0089 \\
    MI\_Pillar\_bake\_Normal           & 0.0059 & 0.0339 & 0.0230 & 0.0218 & 0.0050 \\
    MI\_Pipe\_bake\_Normal             & 0.0202 & 0.1206 & 0.0740 & 0.0693 & 0.0205 \\
    MI\_Roof\_bake\_Normal             & 0.0179 & 0.0514 & 0.0381 & 0.0366 & 0.0082 \\
    MI\_SignalLight\_green\_bake\_Normal & 0.0124 & 0.0544 & 0.0386 & 0.0377 & 0.0074 \\
    MI\_Stairs\_bake\_Normal           & 0.0207 & 0.0784 & 0.0596 & 0.0575 & 0.0168 \\
    MI\_TicketMachine\_bake\_Normal    & 0.0127 & 0.0616 & 0.0465 & 0.0455 & 0.0105 \\
    MI\_Tracks\_bake\_Normal           & 0.0114 & 0.0390 & 0.0258 & 0.0241 & 0.0078 \\
    MI\_TrainParts\_bake\_Normal       & 0.0200 & 0.0749 & 0.0582 & 0.0567 & 0.0133 \\
    MI\_TrainStation\_Floor\_bake\_Normal & 0.0164 & 0.0538 & 0.0432 & 0.0436 & 0.0116 \\
    MI\_Train\_bake\_Normal            & 0.0386 & 0.1231 & 0.0872 & 0.0845 & 0.0318 \\
    MI\_Wall02\_bake\_Normal           & 0.0091 & 0.0594 & 0.0390 & 0.0365 & 0.0050 \\
    MI\_Wall03\_bake\_Normal           & 0.0238 & 0.0777 & 0.0603 & 0.0601 & 0.0148 \\
    MI\_Window\_bake\_Normal           & 0.0273 & 0.1185 & 0.0809 & 0.0770 & 0.0210 \\
    M\_FloorBorderSmall\_bake\_Normal  & 0.0237 & 0.1061 & 0.0798 & 0.0782 & 0.0252 \\
    M\_Ground\_bake\_Normal            & 0.2630 & 0.5785 & 0.5327 & 0.5378 & 0.2864 \\
    M\_Ornaments\_bake\_Normal         & 0.0111 & 0.0546 & 0.0331 & 0.0320 & 0.0062 \\
    M\_Paintings01\_bake\_Normal       & 0.0154 & 0.0593 & 0.0285 & 0.0265 & 0.0065 \\
    M\_Post\_bake\_Normal              & 0.0091 & 0.0412 & 0.0306 & 0.0300 & 0.0069 \\
    M\_Wall01Master\_bake\_Normal      & 0.0098 & 0.0626 & 0.0427 & 0.0403 & 0.0082 \\
    \bottomrule
  \end{tabular}
\end{table*}

\begin{table*}[ht]
  \centering
  \caption{LPIPS per texture in the ARM class in the \textsc{Victorian} dataset}
  \small
  \begin{tabular}{lrrrrr}
    \toprule
    Texture & Ours & Bilinear & Bicubic & Lanczos3 & PFT \\
    \midrule
    MI\_ArchLarge\_bake\_MetallicRoughness        & 0.0367 & 0.0951 & 0.0683 & 0.0668 & 0.0151 \\
    MI\_Bench\_bake\_MetallicRoughness            & 0.0552 & 0.1176 & 0.0975 & 0.0989 & 0.0101 \\
    MI\_Bin\_bake\_MetallicRoughness              & 0.0044 & 0.0156 & 0.0082 & 0.0074 & 0.0027 \\
    MI\_Brick\_bake\_MetallicRoughness            & 0.0343 & 0.1025 & 0.0637 & 0.0633 & 0.0256 \\
    MI\_Ceiling\_bake\_MetallicRoughness          & 0.0671 & 0.1426 & 0.1195 & 0.1205 & 0.0308 \\
    MI\_Clock\_bake\_MetallicRoughness            & 0.0402 & 0.1144 & 0.0753 & 0.0723 & 0.0163 \\
    MI\_ControlGate\_bake\_MetallicRoughness      & 0.0100 & 0.0249 & 0.0152 & 0.0142 & 0.0056 \\
    MI\_Door\_bake\_MetallicRoughness             & 0.0832 & 0.1768 & 0.1593 & 0.1619 & 0.0315 \\
    MI\_FloorBorder\_bake\_MetallicRoughness      & 0.0942 & 0.2044 & 0.1606 & 0.1595 & 0.0519 \\
    MI\_FloorChecker\_bake\_MetallicRoughness     & 0.0883 & 0.1609 & 0.1397 & 0.1406 & 0.0563 \\
    MI\_FloorTileDetails\_bake\_MetallicRoughness & 0.0390 & 0.0996 & 0.0670 & 0.0640 & 0.0142 \\
    MI\_FloorTiles\_bake\_MetallicRoughness       & 0.0061 & 0.0146 & 0.0071 & 0.0065 & 0.0066 \\
    MI\_HalfArch\_bake\_MetallicRoughness         & 0.0374 & 0.1145 & 0.0781 & 0.0757 & 0.0167 \\
    MI\_InfoTable\_bake\_MetallicRoughness        & 0.0274 & 0.0730 & 0.0492 & 0.0468 & 0.0072 \\
    MI\_Light\_bake\_MetallicRoughness            & 0.0416 & 0.0975 & 0.0736 & 0.0726 & 0.0188 \\
    MI\_PhoneBooth\_bake\_MetallicRoughness       & 0.0283 & 0.0800 & 0.0535 & 0.0503 & 0.0122 \\
    MI\_PillarFrame\_bake\_MetallicRoughness      & 0.0787 & 0.1732 & 0.1275 & 0.1263 & 0.0430 \\
    MI\_Pillar\_bake\_MetallicRoughness           & 0.0470 & 0.1348 & 0.0953 & 0.0918 & 0.0216 \\
    MI\_Pipe\_bake\_MetallicRoughness             & 0.0442 & 0.1190 & 0.0808 & 0.0774 & 0.0198 \\
    MI\_Roof\_bake\_MetallicRoughness             & 0.0551 & 0.1521 & 0.0954 & 0.0911 & 0.0261 \\
    MI\_SignalLight\_green\_bake\_MetallicRoughness & 0.0555 & 0.1338 & 0.0968 & 0.0943 & 0.0259 \\
    MI\_Signs\_bake\_MetallicRoughness            & 0.0143 & 0.0513 & 0.0349 & 0.0341 & 0.0046 \\
    MI\_Stairs\_bake\_MetallicRoughness           & 0.0356 & 0.0910 & 0.0672 & 0.0643 & 0.0167 \\
    MI\_TicketMachine\_bake\_MetallicRoughness    & 0.0345 & 0.0910 & 0.0633 & 0.0611 & 0.0122 \\
    MI\_Tracks\_bake\_MetallicRoughness           & 0.0603 & 0.1209 & 0.1038 & 0.1054 & 0.0359 \\
    MI\_TrainParts\_bake\_MetallicRoughness       & 0.0262 & 0.0658 & 0.0479 & 0.0468 & 0.0111 \\
    MI\_TrainStation\_Floor\_bake\_MetallicRoughness & 0.0529 & 0.1335 & 0.0940 & 0.0905 & 0.0272 \\
    MI\_Train\_bake\_MetallicRoughness            & 0.0158 & 0.0480 & 0.0269 & 0.0249 & 0.0083 \\
    MI\_Wall02\_bake\_MetallicRoughness           & 0.0840 & 0.1738 & 0.1341 & 0.1308 & 0.0527 \\
    MI\_Wall03\_bake\_MetallicRoughness           & 0.0690 & 0.1590 & 0.1113 & 0.1106 & 0.0446 \\
    MI\_Window\_bake\_MetallicRoughness           & 0.0314 & 0.0928 & 0.0621 & 0.0585 & 0.0133 \\
    M\_FloorBorderSmall\_bake\_MetallicRoughness  & 0.0671 & 0.1661 & 0.1203 & 0.1160 & 0.0316 \\
    M\_Ground\_bake\_MetallicRoughness            & 0.0427 & 0.1505 & 0.0700 & 0.0663 & 0.0292 \\
    M\_Ornaments\_bake\_MetallicRoughness         & 0.0285 & 0.0967 & 0.0685 & 0.0673 & 0.0109 \\
    M\_Post\_bake\_MetallicRoughness              & 0.0607 & 0.1310 & 0.1055 & 0.1055 & 0.0336 \\
    M\_Wall01Master\_bake\_MetallicRoughness      & 0.0779 & 0.1822 & 0.1293 & 0.1271 & 0.0541 \\
    \bottomrule
  \end{tabular}
\end{table*}

\begin{table*}[ht]
  \centering
  \caption{LPIPS per texture in the diffuse class in the \textsc{Museum} dataset}
  \small
  \begin{tabular}{lrrrrr}
    \toprule
    Texture & baseline & bilinear & bicubic & lanczos3 & PFT \\
    \midrule
    MI\_Banner01\_BaseColor & 0.0090 & 0.0293 & 0.0166 & 0.0157 & 0.0040 \\
    MI\_Bench\_BaseColor & 0.0223 & 0.0658 & 0.0445 & 0.0422 & 0.0169 \\
    MI\_CeilingGlass01\_BaseColor & 0.0848 & 0.2178 & 0.1570 & 0.1508 & 0.0409 \\
    MI\_Cofr\_BaseColor & 0.0414 & 0.1336 & 0.1109 & 0.1133 & 0.0230 \\
    MI\_Column01\_BaseColor & 0.0943 & 0.2463 & 0.2089 & 0.2139 & 0.0961 \\
    MI\_DetailedCeiling01\_BaseColor & 0.0535 & 0.1719 & 0.1344 & 0.1366 & 0.0555 \\
    MI\_DetailedCeiling02\_BaseColor & 0.0482 & 0.1689 & 0.1245 & 0.1237 & 0.0483 \\
    MI\_DetailedCeiling03\_BaseColor & 0.0665 & 0.1918 & 0.1552 & 0.1591 & 0.0701 \\
    MI\_DetailedCeiling04\_BaseColor & 0.0350 & 0.1383 & 0.1014 & 0.0995 & 0.0294 \\
    MI\_Emissive01\_BaseColor & 0.1215 & 0.2223 & 0.1877 & 0.1910 & 0.1359 \\
    MI\_Hebe02\_BaseColor & 0.0653 & 0.1858 & 0.1512 & 0.1554 & 0.0350 \\
    MI\_MetalDoor01\_BaseColor & 0.2322 & 0.3966 & 0.3638 & 0.3643 & 0.2173 \\
    MI\_ModularCeiling\_BaseColor & 0.0172 & 0.0546 & 0.0423 & 0.0419 & 0.0098 \\
    MI\_Painting01\_BaseColor & 0.0381 & 0.1466 & 0.1003 & 0.0934 & 0.0390 \\
    MI\_Painting02\_BaseColor & 0.0376 & 0.1469 & 0.1000 & 0.0912 & 0.0356 \\
    MI\_Painting03\_BaseColor & 0.0266 & 0.1106 & 0.0736 & 0.0668 & 0.0258 \\
    MI\_Painting05\_BaseColor & 0.0286 & 0.1226 & 0.0768 & 0.0711 & 0.0290 \\
    MI\_Painting06\_BaseColor & 0.0302 & 0.1404 & 0.0911 & 0.0847 & 0.0333 \\
    MI\_Painting08\_BaseColor & 0.0206 & 0.0941 & 0.0570 & 0.0518 & 0.0244 \\
    MI\_Painting09\_BaseColor & 0.0178 & 0.0725 & 0.0401 & 0.0356 & 0.0167 \\
    MI\_Painting11\_BaseColor & 0.0445 & 0.1066 & 0.0682 & 0.0664 & 0.0367 \\
    MI\_Painting12\_BaseColor & 0.0497 & 0.1310 & 0.0936 & 0.0958 & 0.0480 \\
    MI\_Painting13\_BaseColor & 0.0385 & 0.1292 & 0.0923 & 0.0956 & 0.0465 \\
    MI\_Painting14\_BaseColor & 0.0422 & 0.1223 & 0.0823 & 0.0820 & 0.0402 \\
    MI\_Painting15\_BaseColor & 0.0635 & 0.1666 & 0.1234 & 0.1270 & 0.0719 \\
    MI\_Painting16\_BaseColor & 0.0138 & 0.0408 & 0.0235 & 0.0229 & 0.0106 \\
    MI\_Painting17\_BaseColor & 0.0640 & 0.1653 & 0.1165 & 0.1152 & 0.0628 \\
    MI\_Painting18\_BaseColor & 0.0471 & 0.1427 & 0.0978 & 0.0954 & 0.0419 \\
    MI\_Painting19\_BaseColor & 0.0577 & 0.1728 & 0.1218 & 0.1190 & 0.0600 \\
    MI\_Painting21\_BaseColor & 0.0369 & 0.1063 & 0.0721 & 0.0703 & 0.0397 \\
    MI\_Painting23\_BaseColor & 0.0268 & 0.0995 & 0.0659 & 0.0652 & 0.0248 \\
    MI\_Painting24\_BaseColor & 0.0154 & 0.0633 & 0.0407 & 0.0424 & 0.0174 \\
    MI\_Painting25\_BaseColor & 0.0320 & 0.0893 & 0.0633 & 0.0651 & 0.0355 \\
    MI\_Painting26\_BaseColor & 0.0483 & 0.1215 & 0.0789 & 0.0781 & 0.0384 \\
    MI\_Painting27\_BaseColor & 0.0487 & 0.1558 & 0.1008 & 0.0973 & 0.0429 \\
    MI\_Painting28\_BaseColor & 0.0374 & 0.1133 & 0.0781 & 0.0797 & 0.0314 \\
    MI\_Painting29\_BaseColor & 0.0371 & 0.1295 & 0.1104 & 0.1160 & 0.0436 \\
    MI\_Painting30\_BaseColor & 0.0363 & 0.0883 & 0.0598 & 0.0606 & 0.0362 \\
    MI\_Painting31\_BaseColor & 0.0290 & 0.0913 & 0.0613 & 0.0605 & 0.0282 \\
    MI\_Philosopher01\_BaseColor & 0.0364 & 0.1556 & 0.1116 & 0.1120 & 0.0402 \\
    MI\_PictureFrames01\_BaseColor & 0.0296 & 0.1002 & 0.0754 & 0.0739 & 0.0177 \\
    MI\_PictureFrames02\_BaseColor & 0.0925 & 0.2060 & 0.1828 & 0.1905 & 0.0745 \\
    MI\_ProjectorLamps\_BaseColor & 0.0295 & 0.0919 & 0.0675 & 0.0650 & 0.0147 \\
    MI\_Railing01\_BaseColor & 0.0496 & 0.1594 & 0.1022 & 0.0949 & 0.0336 \\
    MI\_StageProps\_BaseColor & 0.1299 & 0.2261 & 0.1999 & 0.2013 & 0.0959 \\
    MI\_Staircase01\_BaseColor & 0.1137 & 0.2599 & 0.2261 & 0.2311 & 0.1268 \\
    MI\_Staircase02\_BaseColor & 0.0942 & 0.2212 & 0.1890 & 0.1943 & 0.1019 \\
    MI\_Staircase03\_BaseColor & 0.0956 & 0.2395 & 0.1994 & 0.2024 & 0.0995 \\
    MI\_Staircase04\_BaseColor & 0.0869 & 0.2180 & 0.1837 & 0.1888 & 0.0971 \\
    MI\_StandAndRope\_BaseColor & 0.0036 & 0.0100 & 0.0071 & 0.0071 & 0.0038 \\
    MI\_Stucco01\_BaseColor & 0.0577 & 0.1465 & 0.0992 & 0.0938 & 0.0557 \\
    MI\_WallTrims01\_BaseColor & 0.0791 & 0.2005 & 0.1601 & 0.1620 & 0.0819 \\
    MI\_WallTrims02\_BaseColor & 0.0293 & 0.1088 & 0.0725 & 0.0705 & 0.0203 \\
    MI\_WindowFrame01\_BaseColor & 0.0521 & 0.1128 & 0.0937 & 0.0935 & 0.0322 \\
    MI\_WoodFloor01\_BaseColor & 0.0622 & 0.2052 & 0.1445 & 0.1342 & 0.0595 \\
    M\_Glass01\_StaticMeshActor\_12\_BaseColor & 0.0001 & 0.0000 & 0.0000 & 0.0000 & 0.0006 \\
    \bottomrule
  \end{tabular}
\end{table*}

\begin{table*}[ht]
  \centering
  \caption{LPIPS per texture in the normal class in the \textsc{Museum} dataset}
  \small
  \begin{tabular}{lrrrrr}
    \toprule
    Texture & baseline & bilinear & bicubic & lanczos3 & PFT \\
    \midrule
    MI\_Banner01\_Normal & 0.0351 & 0.0944 & 0.0611 & 0.0569 & 0.0146 \\
    MI\_Bench\_Normal & 0.0338 & 0.1039 & 0.0788 & 0.0756 & 0.0293 \\
    MI\_Cofr\_Normal & 0.0213 & 0.0847 & 0.0693 & 0.0691 & 0.0144 \\
    MI\_Column01\_Normal & 0.0094 & 0.0425 & 0.0285 & 0.0264 & 0.0042 \\
    MI\_DetailedCeiling01\_Normal & 0.0110 & 0.0609 & 0.0422 & 0.0399 & 0.0046 \\
    MI\_DetailedCeiling02\_Normal & 0.0093 & 0.0531 & 0.0360 & 0.0340 & 0.0043 \\
    MI\_DetailedCeiling03\_Normal & 0.0082 & 0.0525 & 0.0381 & 0.0367 & 0.0033 \\
    MI\_DetailedCeiling04\_Normal & 0.0083 & 0.0563 & 0.0357 & 0.0328 & 0.0036 \\
    MI\_FlatColor01\_Normal & 0.0013 & 0.0053 & 0.0039 & 0.0037 & 0.0022 \\
    MI\_Hebe02\_Normal & 0.0160 & 0.0494 & 0.0296 & 0.0277 & 0.0065 \\
    MI\_MetalDoor01\_Normal & 0.0146 & 0.0393 & 0.0274 & 0.0261 & 0.0064 \\
    MI\_ModularCeiling\_Normal & 0.0118 & 0.0406 & 0.0313 & 0.0302 & 0.0081 \\
    MI\_Painting08\_Normal & 0.0479 & 0.0587 & 0.0246 & 0.0219 & 0.0069 \\
    MI\_Philosopher01\_Normal & 0.0164 & 0.0284 & 0.0146 & 0.0124 & 0.0033 \\
    MI\_PictureFrames01\_Normal & 0.0192 & 0.0728 & 0.0538 & 0.0514 & 0.0131 \\
    MI\_PictureFrames02\_Normal & 0.0318 & 0.1064 & 0.0837 & 0.0827 & 0.0226 \\
    MI\_ProjectorLamps\_Normal & 0.0194 & 0.0606 & 0.0456 & 0.0448 & 0.0113 \\
    MI\_Railing01\_Normal & 0.0370 & 0.1220 & 0.0863 & 0.0825 & 0.0216 \\
    MI\_StageProps\_Normal & 0.0570 & 0.1523 & 0.1232 & 0.1241 & 0.0549 \\
    MI\_Staircase02\_Normal & 0.0157 & 0.0351 & 0.0209 & 0.0194 & 0.0041 \\
    MI\_Staircase03\_Normal & 0.0201 & 0.0404 & 0.0244 & 0.0224 & 0.0052 \\
    MI\_Staircase04\_Normal & 0.0188 & 0.0382 & 0.0223 & 0.0206 & 0.0043 \\
    MI\_Staircase05\_Normal & 0.0108 & 0.0253 & 0.0171 & 0.0162 & 0.0048 \\
    MI\_StandAndRope\_Normal & 0.0372 & 0.0760 & 0.0613 & 0.0601 & 0.0276 \\
    MI\_Stucco01\_Normal & 0.0222 & 0.1134 & 0.0683 & 0.0616 & 0.0217 \\
    MI\_WallTrims01\_Normal & 0.0107 & 0.0496 & 0.0376 & 0.0368 & 0.0109 \\
    MI\_WallTrims02\_Normal & 0.0102 & 0.0346 & 0.0232 & 0.0217 & 0.0042 \\
    MI\_WindowFrame01\_Normal & 0.0209 & 0.0729 & 0.0525 & 0.0509 & 0.0122 \\
    MI\_WoodFloor01\_Normal & 0.0058 & 0.0809 & 0.0424 & 0.0362 & 0.0052 \\
    \bottomrule
  \end{tabular}
\end{table*}

\begin{table*}[ht]
  \centering
  \caption{LPIPS per texture in the ARM class in the \textsc{Museum} dataset}
  \small
  \begin{tabular}{lrrrrr}
    \toprule
    Texture & baseline & bilinear & bicubic & lanczos3 & PFT \\
    \midrule
    MI\_Banner01\_Occlusion-MI\_Banner01\_MetallicRoughness & 0.0062 & 0.0257 & 0.0151 & 0.0125 & 0.0030 \\
    MI\_Bench\_Occlusion-MI\_Bench\_MetallicRoughness & 0.0149 & 0.0542 & 0.0293 & 0.0268 & 0.0088 \\
    MI\_CeilingGlass01\_Occlusion-MI\_CeilingGlass01\_MetallicRoughness & 0.0284 & 0.0592 & 0.0410 & 0.0378 & 0.0102 \\
    MI\_Cofr\_Occlusion-MI\_Cofr\_MetallicRoughness & 0.0059 & 0.0139 & 0.0069 & 0.0065 & 0.0016 \\
    MI\_Column01\_Occlusion-MI\_Column01\_MetallicRoughness & 0.0135 & 0.0599 & 0.0324 & 0.0299 & 0.0068 \\
    MI\_DetailedCeiling01\_Occlusion-MI\_DetailedCeiling01\_MetallicRoughness & 0.0387 & 0.1097 & 0.0877 & 0.0881 & 0.0253 \\
    MI\_DetailedCeiling02\_Occlusion-MI\_DetailedCeiling02\_MetallicRoughness & 0.0215 & 0.0786 & 0.0549 & 0.0526 & 0.0125 \\
    MI\_DetailedCeiling03\_Occlusion-MI\_DetailedCeiling03\_MetallicRoughness & 0.0486 & 0.1266 & 0.1047 & 0.1055 & 0.0301 \\
    MI\_DetailedCeiling04\_Occlusion-MI\_DetailedCeiling04\_MetallicRoughness & 0.0244 & 0.0876 & 0.0637 & 0.0621 & 0.0150 \\
    MI\_FlatColor01\_Occlusion-MI\_FlatColor01\_MetallicRoughness & 0.0796 & 0.1427 & 0.1081 & 0.1055 & 0.0489 \\
    MI\_Hebe02\_Occlusion-MI\_Hebe02\_MetallicRoughness & 0.0326 & 0.1085 & 0.0779 & 0.0747 & 0.0222 \\
    MI\_MetalDoor01\_Occlusion-MI\_MetalDoor01\_MetallicRoughness & 0.0356 & 0.0856 & 0.0635 & 0.0615 & 0.0277 \\
    MI\_ModularCeiling\_Occlusion-MI\_ModularCeiling\_MetallicRoughness & 0.0130 & 0.0701 & 0.0320 & 0.0287 & 0.0084 \\
    MI\_Painting08\_MetallicRoughness & 0.0221 & 0.1124 & 0.0655 & 0.0654 & 0.0107 \\
    MI\_Philosopher01\_Occlusion-MI\_Philosopher01\_MetallicRoughness & 0.0169 & 0.0760 & 0.0474 & 0.0440 & 0.0108 \\
    MI\_PictureFrames01\_Occlusion-MI\_PictureFrames01\_MetallicRoughness & 0.0623 & 0.1451 & 0.1165 & 0.1158 & 0.0413 \\
    MI\_PictureFrames02\_Occlusion-MI\_PictureFrames02\_MetallicRoughness & 0.0311 & 0.0917 & 0.0701 & 0.0685 & 0.0259 \\
    MI\_ProjectorLamps\_Occlusion-MI\_ProjectorLamps\_MetallicRoughness & 0.0079 & 0.0249 & 0.0138 & 0.0129 & 0.0029 \\
    MI\_Railing01\_Occlusion-MI\_Railing01\_MetallicRoughness & 0.0225 & 0.0501 & 0.0314 & 0.0292 & 0.0067 \\
    MI\_StageProps\_Occlusion-MI\_StageProps\_MetallicRoughness & 0.0248 & 0.0789 & 0.0487 & 0.0450 & 0.0155 \\
    MI\_Staircase02\_Occlusion-MI\_Staircase02\_MetallicRoughness & 0.0662 & 0.1407 & 0.1253 & 0.1285 & 0.0496 \\
    MI\_Staircase03\_Occlusion-MI\_Staircase03\_MetallicRoughness & 0.0704 & 0.1508 & 0.1350 & 0.1390 & 0.0501 \\
    MI\_Staircase04\_Occlusion-MI\_Staircase04\_MetallicRoughness & 0.0682 & 0.1449 & 0.1293 & 0.1320 & 0.0512 \\
    MI\_Staircase05\_Occlusion-MI\_Staircase05\_MetallicRoughness & 0.0664 & 0.1418 & 0.1251 & 0.1283 & 0.0513 \\
    MI\_StandAndRope\_Occlusion-MI\_StandAndRope\_MetallicRoughness & 0.0116 & 0.0293 & 0.0213 & 0.0205 & 0.0119 \\
    MI\_Stucco01\_Occlusion-MI\_Stucco01\_MetallicRoughness & 0.0142 & 0.0694 & 0.0416 & 0.0417 & 0.0149 \\
    MI\_WallTrims01\_Occlusion-MI\_WallTrims01\_MetallicRoughness & 0.0160 & 0.0512 & 0.0354 & 0.0339 & 0.0095 \\
    MI\_WallTrims02\_Occlusion-MI\_WallTrims02\_MetallicRoughness & 0.0156 & 0.0543 & 0.0325 & 0.0306 & 0.0071 \\
    MI\_WindowFrame01\_Occlusion-MI\_WindowFrame01\_MetallicRoughness & 0.0255 & 0.0768 & 0.0557 & 0.0530 & 0.0192 \\
    MI\_WoodFloor01\_Occlusion-MI\_WoodFloor01\_MetallicRoughness & 0.0264 & 0.1027 & 0.0579 & 0.0527 & 0.0153 \\
    \bottomrule
  \end{tabular}
\end{table*}

\end{document}